\documentclass[aps,prb,twocolumn,superscriptaddress,footinbib,longbibliography]{revtex4-2}

\usepackage{graphicx}
\usepackage{subfigure}
\usepackage{dcolumn}
\usepackage{bm}
\usepackage{times} 
\usepackage{amsmath}
\usepackage{amssymb}
\usepackage{float}
\usepackage{color}
\usepackage{multirow}
\usepackage{url,hyperref}
\usepackage[normalem]{ulem}

\begin{document}
\title{Modulating near-field thermal transfer through temporal drivings: a quantum
many-body theory}

\author{Gaomin Tang}
\email{gmtang@gscaep.ac.cn}
\affiliation{Graduate School of China Academy of Engineering Physics, Beijing 100193,
China}
\author{Jian-Sheng Wang}
\email{phywjs@nus.edu.sg}
\affiliation{Department of Physics, National University of Singapore, Singapore 117551,
Republic of Singapore}

\bigskip

\begin{abstract}
  The traditional approach to studying near-field thermal transfer is based on
  fluctuational electrodynamics. However, this approach may not be suitable for
  nonequilibrium states due to dynamic drivings. In our work, we introduce a theoretical
  framework to describe the phenomenon of near-field heat transfer between two objects
  when subjected to periodic time modulations.
  We utilize the machinery of nonequilibrium Green's function to derive
  general expressions for the DC energy current in Floquet space.
  Furthermore, we also obtain the energy current under the condition of small driving
  amplitude.
  The external drivings create a nonequilibrium state, which gives rise to various effects
  such as heat-transfer enhancement, heat-transfer suppression, and cooling. 
  To illustrate these phenomena, we conduct numerical calculations on a system of
  Coulomb-coupled quantum dots, and specifically investigate the scenario of
  periodically driving electronic reservoir. 
  In our calculations, we employ the $G_0W_0$ approximation, which does not require
  self-consistent iteration and is suitable for weak Coulomb interaction.
  Our theoretical formalism can be applied to study near-field energy transfer
  between two metallic plates under periodic time modulations.
\end{abstract}

\maketitle

\section{Introduction} \label{sec:I}
Near-field heat transfer between objects with nanoscale separations has recently
received significant attention due to its potential in energy harvesting and thermal
management applications~\cite{Pendry99, review07, review15, review15-2, review18,
review20,GT21}. The theoretical treatment is typically formulated within the framework of
fluctuating electrodynamics, which is based on the fluctuation-dissipation theorem by
assuming a thermal equilibrium state for each object~\cite{Rytov, PvH}. This approach
allows for the calculation of the near-field thermal current between objects, taking into
account their geometries, material properties, and the frequency and polarization of the
electromagnetic fields involved~\cite{review20}.

Recently, there has been a growing interest in the use of temporal modulation as a
promising approach to control and manage thermal radiation~\cite{coppens17, Shuttling18,
Kou18, Li19, Fan20, Li21, Li21_ACS, Fan23_1, Fan23_2, Biehs22, Biehs23_1, Biehs23_2,
Lozano23, Picardi23}.
In the near-field regime, Latella et al. reported on the shuttling effect
between two bodies with an oscillating temperature differences or photon chemical
potential difference. However, it is important to note that this work assumes the two
bodies are always in thermal equilibrium~\cite{Shuttling18}. 
Other recent studies have focused on the time modulation of resonance frequencies to generate
synthetic electric and magnetic fields~\cite{Biehs22, Biehs23_1, Biehs23_2}. 
These studies have reported nonreciprocity in the heat transfer transmission function
using quantum Langevin equations. It is worth mentioning that the validity of the
fluctuation-dissipation theorem has been assumed under certain conditions.
Furthermore, the manipulation of spatial coherence in thermal radiation has been achieved
by using a time-modulated lossless layer on top of a semi-infinite lossy
substrate~\cite{Fan23_1, Fan23_2}. The time modulation enables the spatial-coherence
transfer and correlations between different frequency components.
A quantum theoretical formulation, based on macroscopic quantum electrodynamics, has been
proposed by assuming that the time-varying susceptibility modulation is local in
time~\cite{Lozano23}. This theory predicts several nontrivial effects, including nonlocal
correlations between fluctuating currents, far-field thermal radiation surpassing the
black-body spectrum, and quantum vacuum amplification effects.

In addition to the traditional theoretical formalism based on fluctuational
electrodynamics~\cite{PvH}, a fully quantum theoretical framework utilizing the
nonequilibrium Green's function (NEGF) has been established to calculate the near-field
heat current mediated by charge fluctuations~\cite{JSW0, JSW1, GT18, GT19, Wise22, JSW23,
JSW23_2}. 
In this work, we start from the microscopic Hamiltonian and extend the NEGF formalism to
study near-field energy transfer in the presence of periodic time modulations.
We derive general expressions for the energy current in the Floquet space without relying
on the fluctuation-dissipation theorem. Additionally, we expand the energy current up to
the second order of the driving amplitude. 
Using numerical calculations, we demonstrate several effects induced by the
driving, such as heat-transfer enhancement, heat-transfer suppression, and
cooling in a system consisting of Coulomb-coupled quantum dots.

The paper is structured as follows. In Sec.~\ref{sec:II}, the model Hamiltonian and the
derivation of the energy currents from two perspectives are presented.
Some of the details are provided in the appendices. In Sec.~\ref{sec:III}, we discuss the
specific scenario of small driving amplitude. Section~\ref{sec:IV} is devoted to the
numerical results of the system of Coulomb coupled quantum dots, along with their physical
interpretations.  Our work is summarized in Sec.~\ref{sec:V}.

\section{Theoretical Formalism} \label{sec:II}
\subsection{System and Hamiltonian}
\begin{figure}
\centering
\includegraphics[width=2.3in]{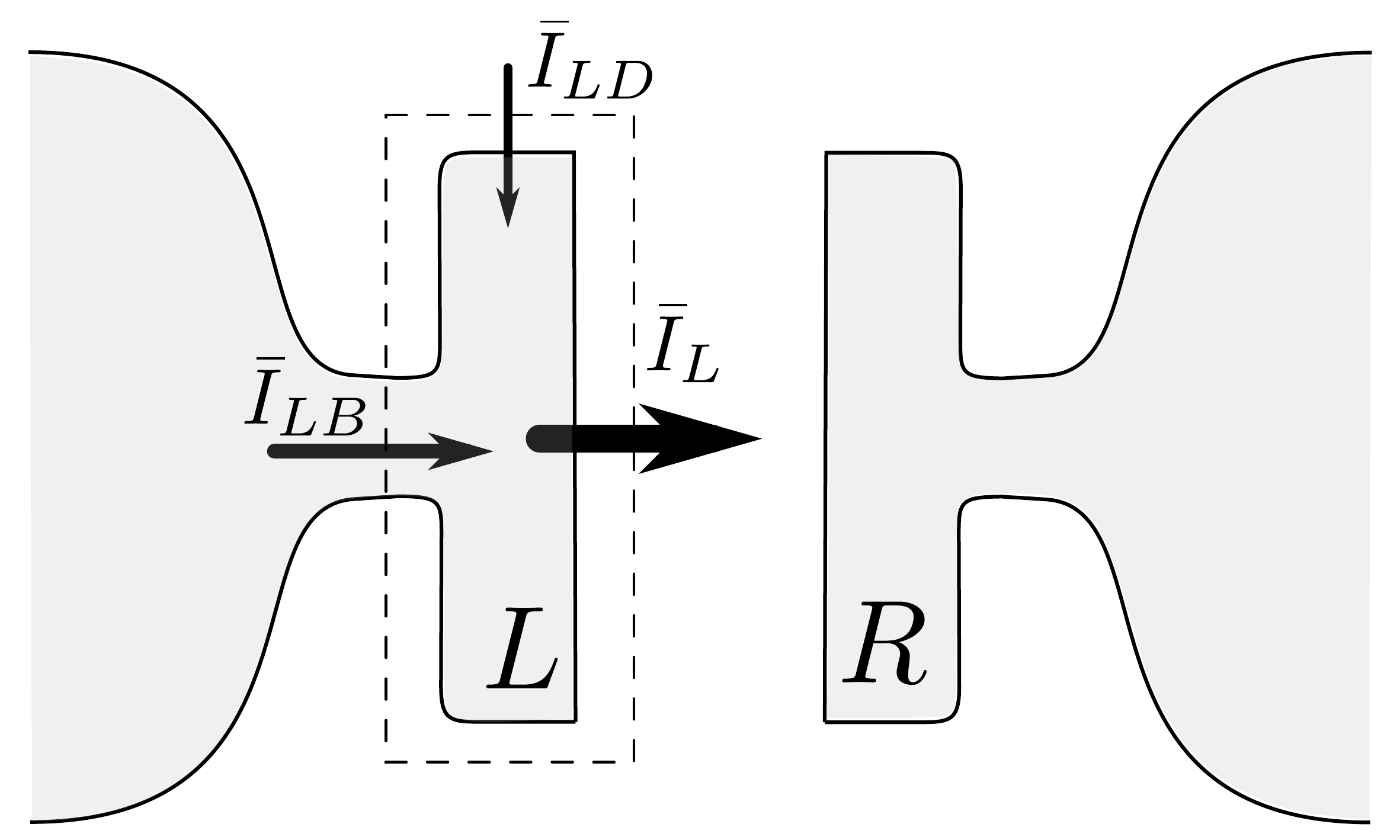} \\
\caption{Schematic plot of the near-field energy transfer between two objects by
  considering charge fluctuations.
  The two subsystems where the electronic reservoirs serving as the heat baths are
  explicitly shown are denoted by $L$ and $R$. The average energy current flowing out of
  heat bath $\alpha$ is denoted as $\bar{I}_{\alpha B}$ with $\alpha=L,R$. 
  With the external driving applied to object $\alpha$ (driving Hamiltonian
  $\hat{H}_\alpha$), the average energy current pumped into the object is $\bar{I}_{\alpha
  D}$. There is a conservation of energy current being expressed as
  $\bar{I}_{\alpha}=\bar{I}_{\alpha B}+\bar{I}_{\alpha D}$. 
  It should be noted that if the driving is only applied to the bath Hamiltonian
  $\hat{H}_{\alpha B}$, we have $\bar{I}_{\alpha}=\bar{I}_{\alpha B}$.
}
\label{fig1}
\end{figure}

We investigate the near-field heat transfer between two objects by considering the
contribution from charge fluctuations, as shown in Fig.~\ref{fig1}. The energy transport
between the two subsystems which are separated by a vacuum gap is mediated by
electromagnetic field. To facilitate energy transport, the electronic reservoirs which
serve as the heat baths in the subsystems are explicitly considered in the theoretical
formalism. This will become clear later when we derive the expression of the energy
current. The total Hamiltonian of the open quantum system can be partitioned as
\begin{equation}
  \hat{H}_{\rm tot} = \hat{H}_e + \hat{H}_{\phi} + \hat{H}_{e \phi} .
\end{equation}
The electronic Hamiltonian is given by
\begin{equation}
  \hat{H}_e = \sum_\alpha \big(\hat{H}_\alpha +\hat{H}_{\alpha B}
  +\hat{V}_{\alpha B}+\hat{V}^\dag_{\alpha B} \big) ,
\end{equation}
where $\alpha=L,R$ denotes the two subsystems [See Fig.~\ref{fig1}]. 
In the tight-binding model, the Hamiltonian of object $\alpha$ is
\begin{equation}
  \hat{H}_\alpha = \sum_{i, j\in \alpha} h_{ij} c_i^\dag c_j ,
\end{equation}
where $i$ and $j$ run over all the electronic sites of object $\alpha$. 
The electronic reservoir which provides the dissipation channel for the energy current
is considered in the Hamiltonian $\hat{H}_{\alpha B}$ with
\begin{equation}
  \hat{H}_{\alpha B} = \sum_{k} \epsilon_{k\alpha} d_{k\alpha}^\dag d_{k\alpha} .
\end{equation}
The coupling between object $\alpha$ and its corresponding reservoir is given by
\begin{equation}
  \hat{V}_{\alpha B} = \sum_{k,i\in\alpha} t_{k\alpha,i} d_{k\alpha}^\dag c_i ,
\end{equation}
which means that an electronic reservoir is attached to every site of the object.
The degrees of freedom of the reservoirs can be analytically integrated out and act as
self-energies in the electronic Green's functions. 
Both $\hat{H}_\alpha$ and $\hat{H}_{\alpha B}$ can be time-dependent, depending on the
problems to study. 
In our numerical calculation in Sec.~\ref{sec:IV}, we drive $\hat{H}_{\alpha B}$ by
applying an AC voltage bias to the electronic reservoir. 
To drive the object Hamiltonian $\hat{H}_\alpha$ via time-dependent electromagnetic
fields, the Peierls substitution which ensures the gauge invariance is used.
This substitution adds a scalar potential to the diagonal term and also introduces a
vector potential as a phase factor in the hopping matrix elements with~\cite{DMFT-RMP14}
\begin{equation}
  h_{ij}(t) = \big[ h_{ij} - e_0 \varphi(\bm{r}_i,t) \delta_{ij} \big] \exp\Big[
  \frac{ie_0}{\hbar} \int_{\bm{r}_i}^{{\bm{r}_j}} d\bm{r} \cdot \bm{A}(\bm{r},t) \Big].
\end{equation}
Here, $\bm{r}_i$ represents the coordinates of electronic site $i$ and $e_0$ denotes the
elementary charge. 
The time-periodic scalar potential $\varphi(\bm{r},t)$ can be achieved by applying a gate
voltage to semiconductor nanostructures, such as quantum dots and quantum wires. In the
case of two-dimensional metallic materials, the scalar potential (or carrier density) can
be dynamically modulated using all-optical techniques~\cite{graphene-modulate-14,
graphene-modulate-19}. Additionally, the time-periodic vector potential $\bm{A}(\bm{r},t)$
can be realized by applying circularly polarized light.

In the context of near-field energy transfer, retardation effects due to the finite speed
of propagation of electromagnetic fields can be disregarded. As a result, we do not need
to consider the vector potential arising from the transverse component of the fluctuating
electric currents, which is responsible for the far-field thermal radiation. Instead, our
focus is on the scalar potential caused by the longitudinal fluctuating electric
currents~\cite{JSW0, Mahan, SPP-graphene3, plasmon20_1, plasmon20_2, Kamenev23,
JSW23}. The observation that the Coulomb interaction is the primary mechanism for
near-field heat transfer between metals was first made by Mahan \cite{Mahan}.

The traditional approach to studying the Coulomb interaction involves eliminating the
scalar field and focusing on the instantaneous interaction. However, in this work, we
consider the scalar field as a fundamental quantum operator and define it in terms of the
usual NEGF method~\cite{JSW0,JSW23}. This approach allows us to treat the Green's
functions for electrons [as shown in Eqs.~\eqref{G_ele_1} and \eqref{G_ele_2}] and the
scalar Coulomb field [as shown in Eq.~\eqref{WK}] on an equal footing. Moreover, it also
facilitates discussing the symmetry properties of the Green's functions [see Appendix C].
The Hamiltonian of the scalar Coulomb field $\phi$ is given by~\cite{QED}
\begin{equation} 
  \hat{H}_{\phi} = - \frac{\epsilon_0}{2} \int d\bm{r} \left[ \big( \dot{\phi}(\bm{r})
  \big)^2 / c_0^2 + \big( \nabla \phi(\bm{r}) \big)^2 \right] .
\end{equation}
Here, $\epsilon_0$ is the vacuum permittivity.  
Due to the instantaneous nature of the Coulomb interaction, the scalar field $\phi$ does
not have a free field dynamics and hence no conjugate momentum from the perspective of
quantum field theory. Therefore, a fictitious speed of light $c_0$ is introduced. 
We take $c_0$ to infinity at the end of the calculations. For further details, we refer to
Ref.~\cite{JSW23}. 
The Hamiltonian that describes the interaction between the Coulomb field and electronic
charges is given by
\begin{equation}
  \hat{H}_{e \phi} = -e_0 \sum_{i} \rho_i \phi(\bm{r}_i) ,
\end{equation}
where $\rho_i = c_i^\dag c_i$ is the number operator at electronic site $i$ with
coordinate $\bm{r}_i$, and $e_0$ represents the elementary charge.
In the limit of $c_0\rightarrow \infty$, this theory is equivalent to the
instantaneous Coulomb problem. 
The Hamiltonian $\hat{H}_{\phi}+\hat{H}_{e \phi}$ captures the electron-electron
interaction expressed by $\sum_{i,j} v_{ij} c_i^\dag c_j^\dag c_j c_i / 2$ with the
interaction strength $v_{ij}=e_0^2/(4\pi\epsilon_0 |\bm{r}_i-\bm{r}_j|)$.

Below we derive the expressions for the DC energy current from both the perspectives of
electronic transport and Coulomb field. 
The equivalence between these two perspectives is proven in Appendix D and Appendix E. 
The main result is presented in Subsection~\ref{sec:II}.C and Section~\ref{sec:III}.
One who is not interested in electronic transport can skip Subsection~\ref{sec:II}.B
without losing any coherence.

\subsection{Perspective from the electronic transport}
We start deriving our formalism from the perspective of electronic transport. 
The average energy current consists of two components: one flowing out of the
electronic reservoir and the other pumped by the external drive [See Fig.~\ref{fig1}].
The energy current flowing out of electronic reservoir $\alpha$ at time $t$ is calculated
in the Heisenberg picture by
\begin{align} \label{IBt}
  I_{\alpha B}(t) =& 
  \frac{1}{i\hbar}\left< \big[ \hat{H}_{\rm tot}(t), \hat{H}_{\alpha B}(t) \big] \right>
  \notag \\
  =& \frac{1}{i\hbar}\left< \big[\hat{V}_{\alpha B}(t)+\hat{V}^\dag_{\alpha B}(t),
  \hat{H}_{\alpha B}(t) \big] \right> \notag \\
  =& \sum_{k,i\in \alpha}\left< t_{k\alpha,i} \dot{d}_{k\alpha}^\dag c_i
    + t^*_{k\alpha,i} c^\dag_i \dot{d}_{k\alpha} \right> .
\end{align}
The ensemble average is defined as $\left< \cdots \right> = \rm{tr}(\hat{\rho}\cdots)$,
where $\hat{\rho}$ is the density-matrix operator. 
The time evolution of $d_{k\alpha}^\dag$ and $d_{k\alpha}$ is governed by the total
Hamiltonian, as shown by 
\begin{align}
  & \big[d_{k\alpha}^\dag, \hat{H}_{\rm tot}\big] = \big[d_{k\alpha}^\dag, \hat{H}_{\alpha
  B}\big] - \sum_i t^*_{k\alpha,i} c^\dag_i , \\
  & \big[d_{k\alpha}, \hat{H}_{\rm tot}\big] = \big[d_{k\alpha}, \hat{H}_{\alpha B}\big] +
  \sum_i t_{k\alpha,i} c_i .
\end{align}
By plugging these equations into Eq.~\eqref{IBt}, the contributions from the second terms
cancel with each other.
It is worth mentioning that the expression of Eq.~\eqref{IBt} in terms of electronic
Green's functions has been derived in many works before~\cite{Haug_Jauho, Chen_2015,
JSW23}. We provide a brief sketch of the derivation for completeness.
By defining the following Green's function on the Keldysh contour with
\begin{equation} \label{G_ele_1}
  G_{i,k\alpha}(\tau,\tau') = \frac{1}{i\hbar} \left< {\cal T} c_i(\tau)
  d^\dag_{k\alpha}(\tau') \right> ,
\end{equation} 
where ${\cal T}$ is the contour time-ordering operator, the thermal current can be
expressed using the lesser component $G^<_{i,k\alpha}$ as
\begin{equation}
  I_{\alpha B}(t) = -i\hbar \sum_{k,i\in \alpha} \Big[t_{k\alpha,i} \partial_{t'}
  G^<_{i,k\alpha}(t,t') \Big]\Big|_{t'=t} + {\rm c.c.} .
\end{equation}
In this work, the time variables in Greek letters sit on the Keldysh contour while those
in Latin letters are normal ones. 
We can express $G_{i,k\alpha}$ in terms of the Green's functions of object $\alpha$
and the corresponding isolated electronic reservoir, $G_{ij}$ and $g_{k\alpha}$,
respectively, as 
\begin{equation} \label{G_ele_2}
  G_{i, k\alpha}(\tau,\tau') = \sum_j\int d\tau_1 G_{ij}(\tau, \tau_1) t^*_{k\alpha,j}
  g_{k\alpha}(\tau_1, \tau') .
\end{equation}
Using the Langreth rule~\cite{Langreth}, the energy current $I_{\alpha B}(t)$ can be
expressed as
\begin{align} 
  I_{\alpha B}(t) = -i\hbar \int_{-\infty}^{\infty} & dt' 
  {\rm tr} \Big[ G_{\alpha}^r(t,t') \partial_t \Sigma_{\alpha}^<(t',t) \notag \\
  + & G_{\alpha}^<(t,t') \partial_t \Sigma_{\alpha}^a(t',t) \Big] + {\rm c.c.} .
\end{align}
Here, the entries of $G_{\alpha}$ and $\Sigma_{\alpha}$ are respectively, $G_{ij}$ and
$\Sigma_{ij}$ with
$\Sigma_{ij}(\tau,\tau') = \sum_{k} t^*_{k\alpha,i} g_{k\alpha}(\tau,\tau')t_{k\alpha,j}$.
The superscripts $<$, $r$, and $a$ denote the lesser, retarded, and advanced components,
respectively. 
In this work, the symbol ${\rm tr}$ denotes trace over electronic sites \textit{only}.
Note that the time derivatives are on the electronic self-energies, and the positions are
important for the driven case. 

We map to the Floquet space, which is explained in more detail in Appendix A.
Using Eqs.~\eqref{F_t2} and \eqref{F_ave}, the DC component $\bar{I}_{\alpha B}\equiv
\int_0^T dt I_{\alpha B}(t) / T$ with period $T = 2\pi/\Omega$ is
\begin{equation} \label{IB}
  \bar{I}_{\alpha B} = \int_{\rm BZ} \frac{dE}{\pi\hbar} {\rm Re}\Big[ {\rm Tr} \Big(
  \bm{E} \bm{G}_{\alpha}^r \bm{\Sigma}_{\alpha}^< + \bm{E} \bm{G}_{\alpha}^<
  \bm{\Sigma}_{\alpha}^a \Big) \Big] ,
\end{equation}
where the symbol ${\rm Tr}$ denotes trace over both the electronic sites and the Floquet
spaces.
It is a generalization of the Meir-Wingreen formula~\cite{Haug_Jauho, Wingreen92} for
the energy transport in Floquet space under periodic modulation. 
Here and below, we use the boldface letters to denote the matrices in Floquet space.
The entries of matrices $\bm{E}$ and $\bm{F}$ are, respectively, $E_m\delta_{m,n}$ and
$F_{mn}$ with $E_m = E+m\hbar\Omega$. 
Note that the integration region is restricted to the ``first Brillouin zone" with
$-\hbar\Omega/2 < E \leq \hbar\Omega/2$, which is denoted as ``BZ". 
Equation~\eqref{IB} gives the energy current flowing out of electronic reservoir $\alpha$.
It is applicable in cases where the electronic reservoir or the object are periodically
driven, or both are driven.

In the scenario where the external drive is applied to object $\alpha$ represented by
$\hat{H}_{\alpha}$, 
the energy current pumped into system $\alpha$ is given by
\begin{equation}
  I_{\alpha D}(t)
  = \sum_{ij\in \alpha}\dot{h}_{ij}(t)\big< c_i^{\dag}(t) c_j(t)\big>
  = -i\hbar \ {\rm tr} \big[ \dot{h}_{\alpha}(t) G_\alpha^<(t,t) \big] .
\end{equation}
Using Eq.~\eqref{AB}, the DC component becomes
\begin{equation}
  \bar{I}_{\alpha D} = \frac{1}{T} \int_0^T i\hbar \ {\rm tr} \Big\{ h_{\alpha}(t)
  \big[ \partial_t G_\alpha^<(t,t) \big] \Big\} dt .
\end{equation}
Using Eqs.~\eqref{F_t1}-\eqref{F_ave}, we obtain its expression in the energy domain as
\begin{equation} \label{ID}
  \bar{I}_{\alpha D} = \int_{\rm BZ} \frac{dE}{2\pi\hbar} {\rm Tr} \Big(
  \bm{h}_{\alpha} \big[\bm{E}, \bm{G}_{\alpha}^< \big] \Big) ,
\end{equation}
where $\bm{h}_{\alpha}$ is a Toeplitz matrix in the Floquet space with entries:
\begin{equation} 
  [\bm{h}_{\alpha}]_{lp} = \frac{1}{T} \int_0^T h_{\alpha}(t) e^{i(l-p)\Omega t} dt .
\end{equation}
The total near-field energy current is the sum of 
$\bar{I}_{\alpha B}$ and $\bar{I}_{\alpha D}$.

\subsection{Perspective from the Coulomb field}
We can also derive the expression of the energy current from the energy density of the
Coulomb field with
\begin{equation}
  u = - \frac{\epsilon_0}{2} \left\{ \big[ \dot{\phi}(\bm{r}) \big]^2 / c_0^2 
    + \big[ \nabla \phi(\bm{r}) \big]^2 \right\} .
\end{equation}
The conservation law in a differential equation form is obtained as~\cite{JSW23}
\begin{align} \label{Poynting}
  \partial_t u
  &= - \epsilon_0 \left[ \dot{\phi}\ddot{\phi}/c_0^2 + \nabla \dot{\phi}
  \cdot \nabla \phi \right] \notag \\
  &= -\nabla \cdot \bm{j} + e_0\sum_i \rho_i \dot{\phi}(\bm{r}_i)\delta(\bm{r}-\bm{r}_i) ,
\end{align}
where we have set $c_0$ to infinity, defined the Poynting vector due to the Coulomb field
as $\bm{j}=\epsilon_0 \dot{\phi}\nabla \phi$, and used the Poisson equation
$\nabla^2\phi(\bm{r}) = e_0 \sum_i \rho_i \delta(\bm{r}-\bm{r}_i) / \epsilon_0$ with
$e_0$ the elementary charge. 
By integrating over volume and using Gauss's law, Eq.~\eqref{Poynting} can be rewritten as 
\begin{equation}
  \oint_{\alpha} \left< \bm{j} \right> \cdot d\bm{A} = 
  e_0 \sum_{i\in \alpha} \big< \rho_i \dot{\phi}(\bm{r}_i) \big> 
  - \int_{\alpha} \big< \partial_t u \big> d\bm{r} ,
\end{equation}
where $\bm{A}$ is the surface that encloses object $\alpha$. The left-hand side describes
the energy current mediated by the Coulomb field. The right-hand side consists of two
terms: the first term describes Joule heating by longitudinal current, while the second
term describes the energy change rate of the dynamical Coulomb field. 
They are respectively denoted as $I_{\alpha}(t)$ and $I_{\alpha d}(t)$. We also refer to
$I_{\alpha d}$ as the displacement energy current. It is analogous to the displacement
current in AC electronic transport~\cite{Buttiker93, Wang99}. 
The displacement energy current only exists in the Coulomb field and does not involve any
real energy transfer, which means that its DC component vanishes, i.e., $\bar{I}_{\alpha
d}=0$. Therefore, the average energy current flowing out of an object is solely
contributed by Joule heating.

In the time domain, the bare (unscreened) Coulomb potential $V$ satisfies 
\begin{equation} \label{V}
  \epsilon_0 \left( \frac{1}{c_0^2} \frac{\partial^2}{\partial t^2} -\nabla^2 \right)
  V(\bm{r}t, \bm{r}'t') = e_0^2 \delta (\bm{r} - \bm{r}') \delta (t-t') .
\end{equation}
It is seen that $e_0^2 V^{-1}=-\epsilon_0\nabla^2$ under $c_0\rightarrow \infty$. 
Using the Poisson equation 
$\nabla^2\phi(\bm{r}) = e_0\sum_i \rho_i \delta(\bm{r}-\bm{r}_i) / \epsilon_0$, we have
\begin{align}
  I_{\alpha}(t)
  =& {\rm Re}\left\{ \epsilon_0 \int_{\alpha} d\bm{r} \left< \big[ \nabla^2\phi(\bm{r} t')
  \big] \dot{\phi}(\bm{r} t) \right> \Big|_{t'=t} \right\}  \notag \\
  =& {\rm Re}\left\{ i\hbar \int_{\alpha} d\bm{r} \big[ V^{-1} \partial_t W^<(\bm{r} t,
  \bm{r} t') \big]\big|_{t'=t} \right\} .
\end{align}
which can be expressed in terms of the dynamically screened Coulomb potential 
defined on the Keldysh contour as 
\begin{equation} \label{WK}
  W(\bm{r}\tau, \bm{r}'\tau') = \frac{e_0^2}{i\hbar} \left< {\cal T}
  \phi(\bm{r}\tau) \phi(\bm{r}'\tau') \right> .
\end{equation}
The dynamically screened Coulomb potential is related to the unscreened Coulomb potential
$V$ and the polarization function $\Pi$ defined in Eq.~\eqref{Pi_K} through the Dyson
equation as shown in Eq.~\eqref{Dyson_W}.
Using Eqs.~\eqref{F_t1} and \eqref{F_ave}, the average of $I_{\alpha}(t)$ in the energy
domain can be expressed as
\begin{align} \label{I}
  \bar{I}_{\alpha} =& {\rm Re}\left[ \int_{\rm BZ} \frac{dE}{2\pi\hbar}
  {\rm Tr}_\alpha \big(\bm{E} \bm{V}^{-1} \bm{W}^< \big) \right] \notag \\
  =& {\rm Re}\left[ \int_{\rm BZ} \frac{dE}{2\pi\hbar} {\rm Tr}_\alpha
  \big( \bm{E} \bm{\Pi}^< \bm{W}^a + \bm{E}\bm{\Pi}^r \bm{W}^< \big) \right] ,
\end{align}
where Eq.~\eqref{W} has been used to get the second identity. 
Equation~\eqref{I} is one of the main results of this work. It is similar in form to
  Eq.~\eqref{IB} where $\bm{W}$ represents the Green's function for the scalar field,
  analogous to the electronic Green's function $\bm{G}$, and the polarization function
  $\bm{\Pi}$ corresponds to self-energy from electronic reservoir. The relation
  $\bar{I}_{\alpha} =\bar{I}_{\alpha B} +\bar{I}_{\alpha D}$ is proven in Appendix D and
  Appendix E. 
Using the symmetries in Eq.~\eqref{sym_GE}, we can alternatively express
$\bar{I}_{\alpha}$ as
\begin{equation} 
  \bar{I}_{\alpha} = \frac{1}{2} \int_{\rm BZ} \frac{dE}{2\pi\hbar}
  {\rm Tr}_{\alpha} \Big[\bm{E} \bm{\Pi}^< \bm{W}^a + \bm{E}\bm{\Pi}^r \bm{W}^<  
  - \big( \bm{\Pi} \leftrightarrow \bm{W} \big)\Big] .
\end{equation}

\section{The case of small driving amplitude} \label{sec:III}
To gain a more intuitive physical understanding, we consider the case where the driving
amplitude is small. By considering up to the second order of the driving amplitude, the
polarization function can be expanded as
\begin{equation} \label{Pi_expand}
  \bm{\Pi}^{\gamma}_{\alpha} =\bm{\Pi}^{\gamma 0}_{\alpha} + \bm{\Pi}^{\gamma 1}_{\alpha}
  + \bm{\Pi}^{\gamma 2}_{\alpha} ,
\end{equation}
with $\gamma = <,r,a$. Here, $\bm{\Pi}^{\gamma 0}_{\alpha}$ represents the polarization
function in the absence of external driving. 
The drivings induce $\bm{\Pi}^{\gamma 1}_{\alpha}$ and $\bm{\Pi}^{\gamma 2}_{\alpha}$,
which are proportional to the first and second order of the driving amplitude,
respectively.
Both $\bm{\Pi}^{\gamma 0}_{\alpha}$ and $\bm{\Pi}^{\gamma 2}_{\alpha}$ are diagonal in
Floquet space. 
On the other hand, the term $\bm{\Pi}^{\gamma 1}_{\alpha}$ is tridiagonal and its value
depends on the phase of the external driving.
Using Eq.~\eqref{W}, the dynamically screened Coulomb potential is thus expanded as
\begin{equation} \label{W_expand}
  \bm{W}^{\gamma} =\bm{W}^{\gamma 0} + \bm{W}^{\gamma 1} + \bm{W}^{\gamma 2} .
\end{equation}
Therefore, in the case of small driving amplitude, the energy current flowing out of the
part $L$ is expressed as
\begin{equation}
  \bar{I}_{L} = \bar{I}_{L 0} + \bar{I}_{L 2} +\bar{I}_{L \theta} .
\end{equation}
The zeroth-order contribution is given by
\begin{equation} 
  \bar{I}_{L 0} = {\rm Re}\bigg\{ \int_{\rm BZ}\frac{dE}{2\pi\hbar}
  {\rm Tr} \big[\bm{E} \big( \bm{\Pi}_L^{<0} \bm{W}_{LL}^{a0} + \bm{\Pi}_L^{r0}
  \bm{W}_{LL}^{<0} \big)\big] \bigg\} ,
\end{equation}
which represents the energy current in the absence of temporal driving.
The term that is first order in driving amplitude vanishes. The second-order contribution
consists of two terms, denoted as $\bar{I}_{L 2}$ and $\bar{I}_{L \theta}$.
The expression of $\bar{I}_{L 2}$ is given by
\begin{align} 
  \bar{I}_{L 2} = {\rm Re}\bigg\{ \int_{\rm BZ}\frac{dE}{2\pi\hbar}
  & {\rm Tr} \big[\bm{E} \big( \widetilde{\bm{\Pi}}_L^{<2} \bm{W}_{LL}^{a0} +
      \widetilde{\bm{\Pi}}_L^{r2} \bm{W}_{LL}^{<0} \notag \\
  + & \bm{\Pi}_L^{<0} \bm{W}_{LL}^{a2} + \bm{\Pi}_L^{r0} \bm{W}_{LL}^{<2} \big)\big]
  \bigg\}, 
\end{align}
with
\begin{equation}
  \widetilde{\bm{\Pi}}_{\alpha}^{r 2} = \bm{\Pi}_{\alpha}^{r 2} + \bm{\Pi}_{\alpha}^{r 1}
  \bm{W}_{\alpha\alpha}^{r 0} \bm{\Pi}_{\alpha}^{r 1}, \quad 
  \widetilde{\bm{\Pi}}_{\alpha}^{a 2}=\Big(\widetilde{\bm{\Pi}}_{\alpha}^{r 2}\Big)^\dag , 
\end{equation}
\begin{equation}
  \widetilde{\bm{\Pi}}_{\alpha}^{< 2} = \bm{\Pi}_{\alpha}^{< 2} + \bm{\Pi}_{\alpha}^{< 1}
  \bm{W}_{\alpha\alpha}^{a 0} \bm{\Pi}_{\alpha}^{a 1} + \bm{\Pi}_{\alpha}^{r 1}
  (\bm{W}_{\alpha\alpha}^{< 0} \bm{\Pi}_{\alpha}^{a 1} + \bm{W}_{\alpha\alpha}^{r 0}
  \bm{\Pi}_{\alpha}^{< 1} ) ,
\end{equation}
\begin{equation}
  \bm{W}_{LL}^{a2} = 
  \bm{W}_{LL}^{a 0} \widetilde{\bm{\Pi}}_L^{a 2} \bm{W}_{LL}^{a 0} +
  \bm{W}_{LR}^{a 0} \widetilde{\bm{\Pi}}_R^{a 2} \bm{W}_{RL}^{a 0} ,
\end{equation}
and
\begin{align}
  \bm{W}_{LL}^{<2} = 
  & \bm{W}_{LL}^{<0} \widetilde{\bm{\Pi}}_L^{a2} \bm{W}_{LL}^{a0} 
  + \bm{W}_{LL}^{r0} \big( \widetilde{\bm{\Pi}}_L^{<2} \bm{W}_{LL}^{a0}
  +  \widetilde{\bm{\Pi}}_L^{r2} \bm{W}_{LL}^{<0} \big) \notag \\
  +& \bm{W}_{LR}^{<0} \widetilde{\bm{\Pi}}_R^{a2} \bm{W}_{RL}^{a0} 
  + \bm{W}_{LR}^{r0} \big( \widetilde{\bm{\Pi}}_R^{<2} \bm{W}_{RL}^{a0}
  + \widetilde{\bm{\Pi}}_R^{r2} \bm{W}_{RL}^{<0} \big) .
\end{align}
It is evident that $\bar{I}_{L 2}$ remains finite as long as one of the electronic
reservoir is subjected to periodic drivings. In the case where both electronic reservoirs
are driven, it is independent of the phase difference between the drivings.
As we will see in the later numerical calculation, this term is relevant to heat-transfer
enhancement, heat-transfer suppression and cooling. 
The expression of $\bar{I}_{L \theta}$ is given by
\begin{equation} 
  \bar{I}_{L \theta} = {\rm Re}\bigg\{ \int_{\rm BZ}\frac{dE}{2\pi\hbar} 
  {\rm Tr} \big[\bm{E} \big( \bm{\Pi}_L^{<1} \widetilde{\bm{W}}_{L}^{a1} + 
  \bm{\Pi}_L^{r1} \widetilde{\bm{W}}_{L}^{<1} \big)\big] \bigg\} ,
\end{equation}
with 
\begin{equation}
  \widetilde{\bm{W}}_{L}^{a1} = \bm{W}_{LR}^{a 0} \bm{\Pi}_R^{a 1} \bm{W}_{RL}^{a 0} ,
\end{equation}
and
\begin{equation}
  \widetilde{\bm{W}}_{L}^{<1} = 
  \bm{W}_{LR}^{<0} \bm{\Pi}_R^{a1} \bm{W}_{RL}^{a0} 
  + \bm{W}_{LR}^{r0} \left( \bm{\Pi}_R^{<1} \bm{W}_{RL}^{a0}
  +  \bm{\Pi}_R^{r1} \bm{W}_{RL}^{<0} \right) .
\end{equation}
The term $\bar{I}_{L \theta}$ is finite only when both reservoirs are driven and depends
on the phase difference between the drivings. 

\section{Numerical Results} \label{sec:IV}
In this section, we present the numerical results for the system of Coulomb-coupled
quantum dots with the Hamiltonian:
\begin{equation}
  \hat{H}_\alpha = \epsilon_\alpha c_\alpha^{\dag} c_\alpha ,
\end{equation}
to demonstrate the physical implications of the periodic temporal modulation. 
We specifically focus on the case of {\it driving the electronic reservoir} with a small
driving amplitude. 
We consider a sinusoidal drive, which can be described by 
$\epsilon_{k\alpha} = \epsilon_{0,k\alpha} + \varepsilon_\alpha(t)$,
where $\varepsilon_\alpha(t) = \mu_\alpha\cos(\Omega t + \theta_\alpha)$ with driving
amplitude $\mu_\alpha$, driving angular frequency $\Omega$ and phase $\theta_\alpha$. 
This driving can be realized through an AC voltage bias applied to the electrode.
The reservoir self-energy $\Sigma_\alpha^\gamma$ with $\gamma = <,r,a$ in the time domain
is given by~\cite{Chen_2015}
\begin{align}
  \Sigma_\alpha^\gamma (t_1, t_2)
  = \Sigma_{\alpha 0}^\gamma (t_1-t_2) \exp\left[ \frac{1}{i\hbar} \int_{t_2}^{t_1}
  \varepsilon_\alpha(t) dt \right] ,
\end{align}
with $\Sigma_{\alpha 0}^\gamma (t_1-t_2)$ representing the self-energy in the absence of
periodic drive and $\kappa_\alpha = \mu_\alpha/(\hbar\Omega)$. 
By using Eq.~\eqref{FmnE_2} and the Jacobi-Anger expansion $\exp(i\kappa\sin x)=\sum_n
J_n(\kappa) \exp(inx)$ with $J_n$ the Bessel function of the first kind, we find
\begin{equation}
  \big[\bm{\Sigma}^\gamma_{\alpha}(E)\big]_{mn} = e^{i(n-m)\theta_\alpha} \sum_{l}
  J_{m-l}(\kappa_\alpha) J_{n-l}(\kappa_\alpha) \Sigma_{\alpha 0}^{\gamma}(E_l). 
\end{equation}
In the wide-band limit, which assumes energy-independent tunneling rates between the quantum
dots
and the electronic reservoirs, we have $\Sigma_{\alpha 0}^{r/a}(E_l)=\mp i\eta$ where $\eta$
is the half linewidth broadened by the electrode. 
Thus, the retarded and advanced self-energies are diagonal in Floquet space with
\begin{equation}
  \big[\bm{\Sigma}^{r/a}_{\alpha}(E)\big]_{mn} =\mp i\eta \delta_{mn} .
\end{equation}
The lesser component $\bm{\Sigma}^{<}_{\alpha}$ is obtained by noticing that
$\Sigma_{\alpha 0}^{<}(E_l) = 2i\eta f_{\alpha,l}$, where the Fermi-Dirac distribution
function is $f_{\alpha,l}=1/\{\exp[\beta_{\alpha}(E_l-\mu_0)]+1\}$
with $\beta_\alpha = 1/(k_B T_\alpha)$. 
We consider the case where the driving amplitude is small, i.e., $\kappa_\alpha \ll 1$.
Expanding to the second order of $\kappa_\alpha$ while maintaining a tridiagonal
structure, one has
\begin{align}
  \big[\bm{\Sigma}^<_{\alpha} & (E)\big]_{mn} = 
  2i\eta \big[ f_{\alpha,m} + (\kappa_\alpha^2/4) (f_{\alpha,m-1} +
  f_{\alpha,m+1}) \big]\delta_{mn}   \notag \\
  & + i\kappa_\alpha\eta (f_{\alpha,m} - f_{\alpha,n})
  \big(e^{i\theta_\alpha} \delta_{m,n-1} -e^{-i\theta_\alpha} \delta_{m,n+1} \big).
\end{align}
The electronic Green's fucntions $\bm{G}_0^r$ and $\bm{G}_0^<$ are obtained through
Eq.~\eqref{G0}.

We use the $G_0W_0$ approximation~\cite{Stefanucci} which requires no self-consistent
iteration and is good for weak Coulomb interaction. 
Having obtained the Green's functions $\bm{G}_0$, the polarization functions and
dynamically screened Coulomb potentials are then calculated using Eqs.~\eqref{Pi} and
\eqref{W}.
In the numerical calculation, the quantum-dot levels are set as equal with $\epsilon_L =
\epsilon_R = 15\,$meV. The half linewidth broadened by the electronic reservoirs is $\eta
= 30\,$meV. The temperatures of the left and right reservoirs are $T_{L(R)} = T_0 \pm
\Delta T /2$ with $T_0=300\,$K. The driving frequency is given by $\hbar\Omega = 5\,$meV.
The Coulomb potential between the two quantum dots is set to be $20\,$meV. 
The parameters of the system under consideration are highly tunable. The quantum-dot
levels and the coupling parameter $\eta$ can be experimentally controlled through gate
voltages. The Coulomb potential is given by $v=e_0^2/(4\pi\epsilon_{\rm eff} d)$, where
$d$ represents the effective distance between the dots and $\epsilon_{\rm eff}$ is the
effective static dielectric constant, which depends on the physical properties of the
quantum dots. It is important to note that the magnitude of the quantum dot levels,
coupling parameter, and Coulomb potential do not influence the physics discussed below.

\begin{figure*}
\centering
\includegraphics[width=\textwidth]{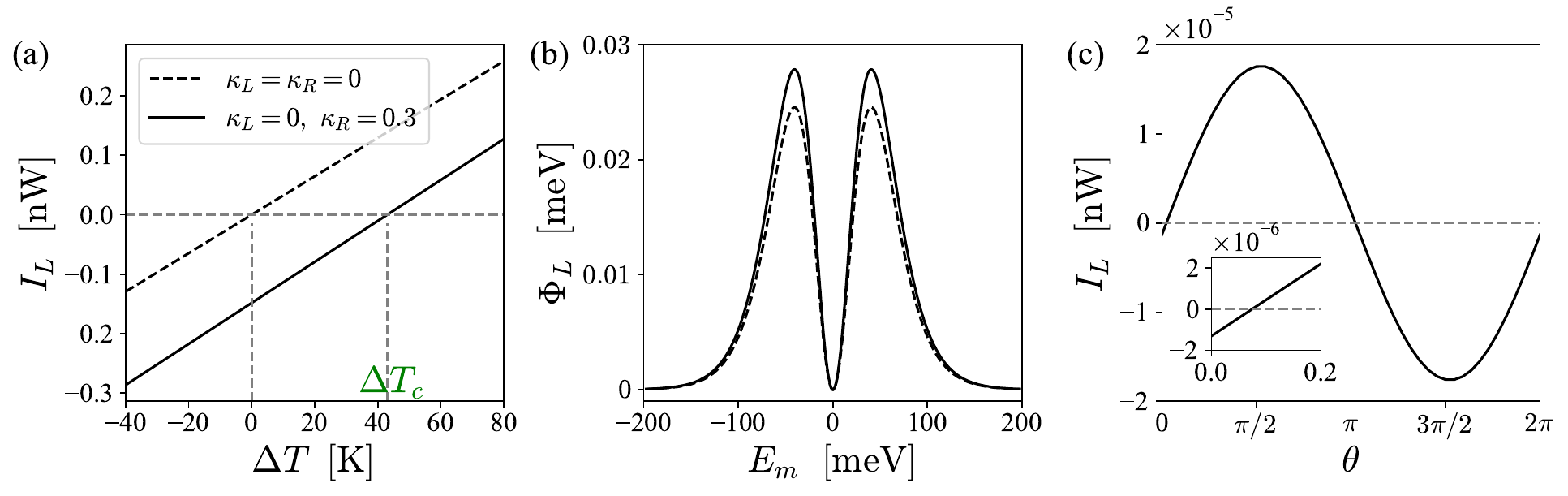} \\
\caption{(a) The energy current $\bar{I}_L$ versus the temperature difference $\Delta
  T\equiv T_L-T_R$, with only electronic reservoir $R$ being driven. 
  The energy current diminishes to zero at a finite $\Delta T$, denoted as $\Delta T_c$.
  The dashed line represents the case without any drivings, for comparison. 
  (b) The spectrum function of the heat current is shown for a temperature difference of
  $\Delta T=40\,$K. The solid line represents the case where both electronic reservoirs
  are subjected to the same periodic driving, while the dashed line represents the case
  without any drivings.
  (c) In the absence of a temperature difference, the energy current versus the phase
  difference $\theta\equiv \theta_L-\theta_R$ is shown when both electronic reservoirs are
  subjected to periodic drivings with the same amplitude and frequency.
  The inset zooms in the curve at small $\theta$. 
}
\label{fig2}
\end{figure*}

We begin by examining the scenario where only the electronic reservoir in part $R$ is
driven with $\kappa_R = 0.3$. Figure~\ref{fig2}(a) illustrates the relationship between
the energy current $\bar{I}_L$ and the temperature difference $\Delta T$. The energy
current diminishes to zero at a certain positive $\Delta T$, referred to as $\Delta T_c$. 
When $\Delta T < 0$ ($T_R > T_L$), the energy current is increased compared to the
undriven case, indicating an enhancement in heat transfer.
In the range of $0 < \Delta T < \Delta T_c$, the energy current continues to flow from
part $R$ to part $L$, suggesting active cooling of the lower temperature region. This
implies that periodic temporal modulation presents a promising alternative for achieving
near-field cooling effect~\cite{photon_potential_15, photon_potential_16,
photon_potential_19,GT21}. 
For $\Delta T > \Delta T_c$, the heat transfer is suppressed due to the periodic driving. 
The unequal energy-current magnitude by reversing the temperature difference allows for
achieving thermal rectification effect. All of these phenomena are a result of the
nonequilibrium state induced by the periodic driving.

We now investigate the scenario in which both electronic reservoirs are subjected to
periodic drivings with the same amplitude and frequency. In the presence of a finite
temperature difference, these drivings enhance the thermal transfer. 
The enhancement is mainly due to $\bar{I}_{L 2}$, and the contribution from $\bar{I}_{L
P}$ can be neglected in this case.
The enhancement can be observed in Fig.~\ref{fig2}(b), which depicts the energy current
spectrum $\Phi_{L}$ defined as
\begin{equation}
  \bar{I}_{L} =\int_{\rm BZ}\frac{dE}{2\pi\hbar} \sum_m \Phi_{L}(E,m) .
\end{equation}
When there is no temperature difference between the two parts, the energy current
$\bar{I}_{L 2}$ becomes zero. However, there is still a small but nonzero energy current
$\bar{I}_{L \theta}$, as shown in Fig.~\ref{fig2}(c). 
Notably, the energy current demonstrates periodicity with respect to the phase difference
of the drivings, denoted as $\theta$. The finite negative energy current at $\theta=0$
suggests that energy is pumped into the system from the drivings.

\section{Conclusion} \label{sec:V}
We have developed a theoretical framework using the nonequilibrium Green's function method
to investigate the near-field energy transport between two bodies when subjected to
periodic temporal drivings. 
The energy current has been derived from both electronic transport and Joule heating, and
the equivalence between these two perspectives has been demonstrated.
In addition, we have expanded the energy current to the second order of the driving
amplitude.
The nonequilibrium state due to the external drivings is responsible for various effects
including heat-transfer enhancement, heat-transfer suppression and cooling.
These have been demonstrated numerically in a system of Coulomb-coupled quantum dots under
$G_0W_0$ approximation. Our formalism can be extended to study more complex systems, such
as the near-field energy transfer between two metallic plates where the electron densities
are dynamially modulated.

\begin{acknowledgments}
G.T. is supported by  National Natural Science Foundation of China (Grant No. 12374048,
and No. 12088101) and NSAF (Grant No. U2330401). 
J.-S.W. acknowledges support from MOE FRC tier 1 grant A-8000990-00-00. 
\end{acknowledgments}

\appendix
\numberwithin{equation}{section}

\section{Floquet representation}
\renewcommand{\theequation}{A.\arabic{equation}}
We give a brief introduction to the Floquet representation of a function $F$ that has two
independent arguments of time, $t_1$ and $t_2$~\cite{Tsuji08, DMFT-RMP14, Kosov23_1,
Kosov23_2}. 
By defining variables $t_r=t_1-t_2$ and $t_a=(t_1+t_2)/2$, it is also possible to write in
its Wigner form, $F^w$, as follows:
\begin{equation}
  F(t_1, t_2) \equiv F^w(t_a, t_r) .
\end{equation}
In equilibrium, $F$ only depends on $t_r$ because of continuous-time translational
invariance. However, in the presence of a periodic drive, the system is out of
equilibrium, causing $F$ to depend on both $t_r$ and $t_a$.
Maintained by the discrete-time translational invariance, we have the following
periodicity,
\begin{equation}
  F(t_1+T, t_2+T) = F(t_1, t_2) , \ \
  F^w(t_a+T, t_r) = F^w(t_a, t_r) ,
\end{equation}
where $T$ is the period of the driving field. 

The Wigner transformation of $F^w(t_a, t_r)$ into energy domain with energy $E$ and single
integer index $l$ is given by
\begin{equation} \label{barF_l}
  F^w_l(E) = \frac{1}{T} \int_0^T dt_a \int_{-\infty}^{\infty} dt_r
  e^{i(E/\hbar) t_r + il\Omega t_a} F^w(t_a, t_r) ,
\end{equation}
where $\Omega = 2\pi /T$ is the driving angular frequency. 
The inverse transformation is
\begin{equation} \label{barF}
  F^w(t_a, t_r) = \sum_{l\in \mathbb{Z}} \int_{-\infty}^{\infty}\frac{dE}{2\pi\hbar}
  e^{-i(E/\hbar) t_r - il\Omega t_a} F^w_l(E) .
\end{equation}
The Floquet representation of $F$ with Floquet indices $m$ and $n$ can be related to the
Wigner representation via
\begin{equation} \label{F-W}
  F_{mn}(E) \equiv F^w_{m-n}\Big( E+\frac{m+n}{2}\hbar\Omega \Big) .
\end{equation}
The range of $E$ in $F_{mn}(E)$ is limited to the ``first Brillouin zone" (BZ), which is
$-\hbar\Omega/2 < E \leq \hbar\Omega/2$~\cite{Tsuji08,DMFT-RMP14}. 
As a result, Eq.~\eqref{barF_l} is equivalent to
\begin{align} \label{FmnE}
  & F_{mn}(E) = 
  \notag \\
  & \frac{1}{T} \int_0^T dt_a \int_{-\infty}^{\infty} & dt_r \ F(t_1,t_2)
  e^{i\big(\frac{E}{\hbar}+\frac{m+n}{2}\Omega\big) t_r + i(m-n)\Omega t_a} .
\end{align}
To simplify this expression, we assume $0<t_2\leq T$ and $t_1=lT+t_1'$ with integer $l$,
so that $0<t_1'\leq T$. The exponential phase factor in Eq.~\eqref{FmnE} is given by
$E_m(t_1'+lT)/\hbar-E_nt_2/\hbar$ with $E_m = E+m\hbar\Omega$ and $E_n = E+n\hbar\Omega$.
Using the Jacobian
\begin{equation}
  \frac{\partial(t_r, t_a)}{\partial(t_1',t_2)} =1,
\end{equation}
Eq.~\eqref{FmnE} can be rewritten as
\begin{align} \label{FmnE_2}
  F_{mn}(E) = \frac{1}{T} \int_0^T dt_2 \int_{-\infty}^{\infty} dt_1 F(t_1,t_2)
  e^{iE_mt_1/\hbar-iE_nt_2/\hbar}.
\end{align}
Equation~\eqref{barF} is equivalent to
\begin{equation}
  F(t_1, t_2) = \sum_{l,p} \int_{\rm BZ} \frac{dE}{2\pi\hbar} 
  e^{-i(E_p/\hbar) t_r - il\Omega t_a} F^w_l(E_p) ,
\end{equation}
where the integral interval denoted by ``BZ" is $-\hbar\Omega/2 < E \leq \hbar\Omega/2$.
Changing the variables $l$ and $p$ to $m$ and $n$ through the transformations $l=m-n$ and
$p=(m+n)/2$, we arrive at a more symmetric form with
\begin{equation} \label{FtFE}
  F(t_1,t_2) = \sum_{mn} \int_{\rm BZ} \frac{dE}{2\pi\hbar} 
  F_{mn}(E) e^{-iE_m t_1/\hbar} e^{iE_n t_2/\hbar} .
\end{equation}
The diagonal elements represent the dependence on the time difference $t_1-t_2$, while
the off-diagonal elements represent the dependence on the average time $(t_1+t_2)/2$.
The derivatives with respect to time indices are 
\begin{align} 
  & \partial_{t_1} F(t_1,t_2) = -\frac{i}{\hbar} \sum_{mn} \int_{\rm BZ}
  \frac{dE}{2\pi\hbar} (\bm{E F})_{mn} e^{-i(E_m t_1-E_n t_2)/\hbar} , \label{F_t1} \\
  & \partial_{t_2} F(t_1,t_2) = \frac{i}{\hbar} \sum_{mn} \int_{\rm BZ}
  \frac{dE}{2\pi\hbar} (\bm{F E})_{mn} e^{-i(E_m t_1-E_n t_2)/\hbar} , \label{F_t2}
\end{align}
where $\bm{E}$ and $\bm{F}$ are matrices with entries $E_m\delta_{m,n}$ and $F_{mn}$,
respectively.
For $t_1=t_2 = t$, the time average of $F(t,t)$ can be obtained as
\begin{equation} \label{F_ave}
  \bar{F} \equiv \frac{1}{T} \int_0^T F(t,t) dt 
  =\sum_{n} \int_{\rm BZ} \frac{dE}{2\pi\hbar} F_{nn}(E) . 
\end{equation}

To provide a more complete understanding, let us relate $F_{mn}(E)$ to $F(E,E')$ which is
defined as
\begin{equation}
  F(E,E') = \iint dt_1 dt_2 e^{i E t_1/\hbar - i E' t_2/\hbar} F(t_1,t_2) .
\end{equation}
By substituting Eq.~\eqref{FtFE} into the above equation, we obtain
\begin{equation}
  F(E,E') = \sum_{mn} 2\pi\hbar \delta \big( E-E'+(n-m)\hbar\Omega \big)
  F_{mn}(E-m\hbar\Omega) ,
\end{equation}
where $m$ is fixed by restricting $E-m\hbar\Omega$ to the range of $(-\hbar\Omega/2,
\hbar\Omega/2]$. This equation demonstrates that $F(E,E')$ is nonzero only when $E$ and
$E'$ differ by a multiple of $\hbar\Omega$.

The Floquet representation preserves the multiplication structure by mapping from the time
domain to the energy domain with
\begin{align} \label{CAB}
  & C(t,t') = \int dt'' A(t,t'') B(t'',t') \notag \\
  \rightarrow \ \
  & \bm{C}(E) = \bm{A}(E) \bm{B}(E) .
\end{align}
We also have the following rules:
\begin{align} \label{CAB_1}
  & C(t_1,t_2) = A(t_1,t_2) B(t_1,t_2) \notag \\
  \rightarrow \ \
  & \bm{C}_{mn}(E) 
  = \sum_{l p}\int_{\rm BZ} \frac{dE'}{2\pi\hbar} \bm{A}_{l,p}(E') \bm{B}_{m-l,n-p}(E-E'),
\end{align}
and
\begin{align} \label{CAB_2}
  & C(t_1,t_2) = A(t_1,t_2) B(t_2,t_1) \notag \\
  \rightarrow \ \
  & \bm{C}_{mn}(E) 
  = \sum_{l p}\int_{\rm BZ} \frac{dE'}{2\pi\hbar} \bm{A}_{l,p}(E') \bm{B}_{p-n,l-m}(E'-E).
\end{align}
The mappings in Eqs.~\eqref{CAB_1} and \eqref{CAB_2} are, respectively, useful for
obtaining the electronic self-energies due to Coulomb interactions and the polarization
functions in energy space.

\section{Nonequilibrium Green's function (NEGF)}
\renewcommand{\theequation}{B.\arabic{equation}}
The Dyson equation on the Keldysh contour, which does not take into account Coulomb
interactions, can be expressed as 
\begin{align} \label{Dyson_G0}
  & G_0(\tau,\tau') = \notag \\
  & g(\tau,\tau') + \iint d\tau_1 d\tau_2 g(\tau,\tau_1)
  \Sigma_B(\tau_1,\tau_2) G_0(\tau_2,\tau') ,
\end{align}
where $g$ represents the Green's function of the free electronic system, and $\Sigma_B$
represents the self-energies due to the coupling to the electronic reservoirs. 
All time variables are defined on the Keldysh contour. The free electronic Green's
function is given by $g_{\alpha}^{-1} = i\hbar \partial_\tau - h_\alpha(\tau)$.
Electronic sites are considered using matrix multiplication in Eq.~\eqref{Dyson_G0}.
Including Coulomb interactions with the Hartree self-energy $\Sigma_H$ and
Fock-like self-energy $\Sigma_\phi$, we have the following expression:
\begin{equation} \label{Dyson_G}
  G = G_0 + G_0 (\Sigma_\phi + \Sigma_H) G ,
\end{equation}
where the time variables and the integral as in Eq.~\eqref{Dyson_G0} have been omitted.
The structures of $g$, $\Sigma_B$, $G_0$, $\Sigma_\phi$, $\Sigma_H$, and $G$ are all
diagonal in subsystem space. Taking $G$ as an example, it can be represented as a block
matrix as
\begin{equation}
  G = 
  \begin{bmatrix}
    G_{L} & 0 \\ 0 & G_{R}
  \end{bmatrix} .
\end{equation}
Under $GW$ approximation~\cite{Stefanucci}, the self-energy $\Sigma_{\phi}$ has the form 
\begin{equation} \label{GW}
  \Sigma_{\phi}(\tau,\tau') = i\hbar G(\tau,\tau') W(\tau,\tau') ,
\end{equation}
where $W$ is the dynamically screened Coulomb potential. 
The explicit expressions for the components of $\Sigma_{\phi}$ can be obtained using
Langreth's theorem~\cite{Langreth, Haug_Jauho} as
\begin{align}
  & \Sigma_{\phi}^<(1,2) = i\hbar G^<(1,2)W^<(1,2) , \label{GW_<_t} \\
  & \Sigma_{\phi}^>(1,2) = i\hbar G^>(1,2)W^>(1,2) , \label{GW_>_t} \\
  & \Sigma_{\phi}^r(1,2) = i\hbar \big[ G^r(1,2)W^<(1,2) +G^>(1,2)W^r(1,2) \big], 
  \label{GW_r_t} \\
  & \Sigma_{\phi}^a(1,2) = i\hbar \big[ G^a(1,2)W^<(1,2) +G^>(1,2)W^a(1,2) \big],
  \label{GW_a_t}
\end{align}
where the indices $1$ and $2$ include those of time and electronic sites. 
The lesser and greater components of the Hartree self-energy vanish. The retarded and
advanced components are equal and diagonal in electronic site space. They are expressed as
\begin{align}
  & \Sigma_{H,jj}^r(t,t') = \Sigma_{H,jj}^a(t,t') = \notag \\
  & -i\hbar \delta(t-t') \sum_k \int dt_1
  V(\bm{r}_j t,\bm{r}_k t_1) G_{kk}^<(t_1,t_1) .
\end{align}
The Hartree term for homogeneous systems is cancelled out by the contribution from the
background ions. 
We will demonstrate later on that it does not contribute to the energy transport. 

In the time domain, the bare Coulomb potential satisfies
\begin{equation}
  \epsilon_0 \left( \frac{1}{c_0^2} \frac{\partial^2}{\partial t^2} -\nabla^2 \right)
  V(\bm{r}t, \bm{r}'t') = e_0^2 \delta (\bm{r} - \bm{r}') \delta (t-t') .
\end{equation}
Under $c_0\rightarrow \infty$, we ignore the term of $\partial^2 / \partial t^2$.
The retarded and advanced compoents of $V$ are equal and given by~\cite{JSW23} 
\begin{equation}
  V(\bm{r}t,\bm{r}'t')  = \frac{e_0^2}{4\pi\epsilon_0 |\bm{r}-\bm{r}'|} \delta(t-t') .
\end{equation}
The instantaneous nature results in the vanishing of its lesser and greater components.
The Dyson equation for the screened Coulomb potential $W$ on the Keldysh contour can be
expressed as
\begin{align} \label{Dyson_W}
  & W(\bm{r}\tau,\bm{r}'\tau') = V(\bm{r}\tau,\bm{r}'\tau') +
  \notag \\
  & \sum_{ij} \iint d\tau_1 d\tau_2 V(\bm{r}\tau,\bm{r}_i\tau_1) 
  \Pi(\bm{r}_i\tau_1,\bm{r}_j\tau_2) W(\bm{r}_j\tau_2,\bm{r}'\tau') .
\end{align}
The polarization function on the Keldysh contour is defined as
\begin{equation} \label{Pi_K}
  \Pi(\bm{r}_i\tau, \bm{r}_j\tau') = \frac{1}{i\hbar} \left< {\cal T} \rho_i(\tau)
  \rho_j(\tau') \right> . 
\end{equation}
Under the random phase approximation, we have
\begin{equation} \label{GG}
  \Pi(\bm{r}_i\tau, \bm{r}_j\tau') = -i\hbar G_{ij}(\tau,\tau') G_{ji}(\tau',\tau) .
\end{equation}
Here, only the irreducible diagrams are considered in a Feynman-diagrammatic expansion
with the Coulomb interaction in defining the polarization function.
Its components are obtained using Langreth theorem as
\begin{align}
  & \Pi^<(1,2) = -i\hbar G^<(1,2) G^>(2,1) \label{GG_<_t} , \\
  & \Pi^>(1,2) = -i\hbar G^>(1,2) G^<(2,1) \label{GG_>_t} , \\
  & \Pi^r(1,2) = -i\hbar \big[ G^r(1,2) G^<(2,1) + G^<(1,2) G^a(2,1) \big]
  \label{GG_r_t} , \\
  & \Pi^a(1,2) = -i\hbar \big[ G^a(1,2) G^<(2,1) + G^<(1,2) G^r(2,1) \big]
  \label{GG_a_t} .
\end{align}
As the electronic Green's function $G$, $\Pi$ is diagonal in subsystem space with
\begin{equation} \label{Pi_L_R}
  \Pi = 
  \begin{bmatrix}
    \Pi_L & 0 \\ 0 & \Pi_R 
  \end{bmatrix} .
\end{equation}
Since $\Pi$ is only nonzero on electronic sites, the relevant coordinates of $W$
in calculating energy current are those of the electrons, as shown in Eq.~\eqref{I}. 
Therefore, it is convenient to partition $V$ and $W$ in numerical calculation using
subsystem space as shown below with
\begin{equation}
  V = 
  \begin{bmatrix}
    V_{LL} & V_{LR} \\ V_{RL} & V_{RR}
  \end{bmatrix} , \qquad 
  W = 
  \begin{bmatrix}
    W_{LL} & W_{LR} \\ W_{RL} & W_{RR}
  \end{bmatrix} .
\end{equation}

Under the $GW$ approximation, it is necessary to calculate $G$, $\Sigma_\phi$, $W$, and
$\Pi$ self-consistently, which can be a challenging task involving the Floquet space. To
reduce computational costs, we take the $G_0W_0$ approximation, which does not require
self-consistent iteration and is valid for weak Coulomb interactions. In this approach, we
use $G_0$ instead of $G$ to obtain the polarization functions $\Pi$ through
Eqs.~\eqref{GG_<_t}-\eqref{GG_a_t}. We then obtain the dynamically screened Coulomb
potential $W$. Furthermore, we obtain the interaction-induced self-energy $\Sigma_\phi$ by
replacing $G$ with $G_0$ in Eqs.~\eqref{GW_<_t}-\eqref{GW_a_t}. The components of the
electronic Green's function used to calculate $\bar{I}{\alpha B}$ and $\bar{I}_{\alpha D}$
are obtained via Eqs.~\eqref{Dyson_G}.

In the energy domain, the Dyson and the Keldysh equations for the electronic Green's
function $\bm{G}_0$ are, respectively,
\begin{equation} \label{G0}
  \bm{G}_0^r = \big(\bm{E} - \bm{h} - \bm{\Sigma}_B^r \big)^{-1} , \quad
  \bm{G}_0^< = \bm{G}_0^r \bm{\Sigma}_{B}^< \bm{G}_0^a .
\end{equation}
To obtain the Floquet representation of the polarization function, we apply
Eq.~\eqref{CAB_2} to Eqs.~\eqref{GG_<_t}-\eqref{GG_a_t}. For example, considering the
lesser component, we have the following equation under the $G_0W_0$ approximation:
\begin{align} \label{Pi}
  \bm{\Pi}^<_{mn}(i,j, E) = -i
  & \int_{\rm BZ} \frac{dE'}{2\pi} \sum_{l p} [\bm{G}_0^<]_{l+m,p+n}(i,j, E')  \notag \\
  & \times [\bm{G}_0^>]_{p,l}(j,i, E'-E) .
\end{align}
where $i$ and $j$ in the brackets denote the electronic site indices.
For the dynamically screened Coulomb potential, we have
\begin{equation} \label{W}
  \bm{W}^r = \bm{V} + \bm{V} \bm{\Pi}^r \bm{W}^r , \quad
  \bm{W}^< = \bm{W}^r \bm{\Pi}^< \bm{W}^a .
\end{equation}
Given $\bm{\Pi}$ and $\bm{W}$, the energy current can be calculated from Eq.~\eqref{I}. 

We apply Eq.~\eqref{CAB_1} to Eqs.~\eqref{GW_<_t}-\eqref{GW_a_t} to obtain the Floquet
representation of the Fock-like self-energy $\Sigma_{\phi}$. Taking the lesser component
as an example, we have
\begin{align}
  \bm{\Sigma}^<_{\phi, mn}(i,j, E) = i
  & \int_{\rm BZ} \frac{dE'}{2\pi} \sum_{l p} [\bm{G}_0^<]_{l,p}(i,j, E')  \notag \\
  & \times [\bm{W}_0^<]_{m-l,n-p}(i,j, E-E') .
\end{align}
The retarded and advanced components of the Hartree self-energy are expressed as
\begin{align}
  & \bm{\Sigma}^r_{H, mn}(j,j, E) = \bm{\Sigma}^a_{H, mn}(j,j, E) = \notag \\
  & -i\sum_k V(j,k) \int_{\rm BZ} \frac{dE'}{2\pi} 
   \sum_{l} [\bm{G}_0^<]_{l+m,l+n}(k,k, E') .
\end{align}
The electronic Green's functions $\bm{G}$ in energy domain can then be obtained through
Eq.~\eqref{Dyson_G} with the Dyson and Keldysh equations, respectively, as
\begin{align}
  & \bm{G}^r = \bm{G}_0^r +\bm{G}_0^r(\bm{\Sigma}^r_\phi +\bm{\Sigma}^r_H) \bm{G}^r , \\
  & \bm{G}^< = \bm{G}^r (\bm{\Sigma}^<_\phi + \bm{\Sigma}_{B}^<) \bm{G}^a .
\end{align}
The information of the nonequilibrium electron distributions is contained in $\bm{G}^< $.

\section{Symmetries of the Green's functions}
\renewcommand{\theequation}{C.\arabic{equation}}
Starting from the basic definition of the electronic Green's function, we have its
symmetries in the time domain as 
\begin{equation} \label{sym_G}
  [G^r(2,1)]^* =  G^a(1,2) , \quad
  [G^{<,>}(2,1)]^* = -G^{<,>}(1,2) ,
\end{equation}
where indices $1$ and $2$ include both time and electronic sites.
Moving on to the energy domain with Floquet representation, one has
\begin{equation} \label{sym_GE}
  [\bm{G}^r(E)]^\dag = \bm{G}^a(E) , \
  [\bm{G}^<(E)]^\dag = -\bm{G}^<(E) ,
\end{equation}
where the Hermitian conjugate operator $\dag$ swaps both Floquet and electronic indices
and then takes complex conjugate. The symmetries of $\bm{G}$ remain consistent for
$\bm{\Sigma}_{\alpha}$, $\bm{\Sigma}_{\phi}$, $\bm{W}$, and $\bm{\Pi}$. 

Additionally, using the bosonic commutation relation, we have further symmetries in the
time domain for the dynamically screened Coulomb potential:
\begin{equation} \label{sym_W}
  W^r(2,1) =  W^a(1,2) , \quad W^>(2,1) =  W^<(1,2) .
\end{equation}
In the energy domain, these symmetries can be expressed as:
\begin{align} 
  & \bm{W}^r_{m,n}(i,j, E) = \bm{W}^a_{-n,-m}(j,i, -E) ,
  \label{sym_W_E1} \\
  & \bm{W}^>_{m,n}(i,j, E) = \bm{W}^<_{-n,-m}(j,i, -E) ,
  \label{sym_W_E2}
\end{align}
where $m$ and $n$ are Floquet indices, and the electronic sites are indicated by $i$ and $j$. 
The symmetries of $\bm{W}$ are also applicable for $\bm{\Pi}$.

\section{Proof of energy conservation on the operator level}
\renewcommand{\theequation}{D.\arabic{equation}}
We demonstrate the equivalence between the perspectives from electronic transport and
Coulomb field by showing the conservation of average energy currents with
\begin{equation} \label{conservation}
  \bar{I}_{\alpha} = \bar{I}_{\alpha B} + \bar{I}_{\alpha D} ,
\end{equation}
on the operator level. 
Equation~\eqref{conservation} implies that the average energy current emitted by object
$\alpha$, $I_{\alpha}$, includes the average energy current taken from the electronic
reservoir, as well as that pumped in by external drive. Let us define the Hamiltonian of a
finite subsystem as:
\begin{equation}
  \hat{H}_{\alpha C} = \hat{H}_{\alpha} + \hat{V}_{\alpha\phi} .
\end{equation}
The energy current that flows out of this subsystem can be calculated as:
\begin{align}
  I_{\alpha C}(t) 
  =& -\left< d_t \hat{H}_{\alpha C} \right>  \notag \\
  =& \frac{1}{i\hbar} \left< \big[ \hat{H}_{\rm tot}, \hat{H}_{\alpha C} \big] \right> 
  - \left< \partial_t \hat{H}_{\alpha C} \right>  \notag \\
  =& \frac{1}{i\hbar} \left< \big[ \hat{V}_{\alpha B} +\hat{V}^\dag_{\alpha B}
  +\hat{H}_{\phi}, \hat{H}_{\alpha C} \big] \right> - I_{\alpha D}(t) .
\end{align}
Similarly to how we obtained the last equality in Eq.~\eqref{IBt}, we have
\begin{equation}
  \frac{1}{i\hbar} \big[ \hat{V}_{\alpha B} +\hat{V}^\dag_{\alpha B},
  \hat{H}_{\alpha C} \big]
  = \sum_{k,i\in \alpha} t_{k\alpha,i} d_{k\alpha}^\dag \dot{c}_i + t^*_{k\alpha,i}
  \dot{c}_i^\dag d_{k\alpha} ,
\end{equation}
where time derivatives in $\dot{c}_i^\dag$ and $\dot{c}_i$ denote the time evolution with
respect to the total Hamiltonian. 

For a periodically driven system, the ensemble average of an operator $\hat{A}$ with one
time variable is periodic with $\big< \hat{A}(t) \big> = \big< \hat{A}(t+T) \big>$, where
$T$ is the driving period, so that the time derivative of the ensemble average is in
general finite with
$\partial_t \big< \hat{A}(t) \big> = \big< \partial_t \hat{A}(t) \big> \neq 0$.
By further taking the time average, which is denoted by a bar, one has
\begin{equation} \label{avg}
  \overline{\big< \partial_t \hat{A}(t) \big>} = 0 .
\end{equation}
This property implies that
\begin{equation}  \label{AB}
  \overline{\left< \partial_t (\hat{A}\hat{B}) \right>} = 
  \overline{\left< \dot{\hat{A}} \hat{B} \right>} + 
  \overline{\left< \hat{A} \dot{\hat{B}} \right>} =0 .
\end{equation}

Using Eq.~\eqref{AB}, we obtain 
\begin{equation}
  \overline{\left< t_{k\alpha,i} d_{k\alpha}^\dag \dot{c}_i + {\rm H.c.} \right>}
  = -\overline{\left< t_{k\alpha,i} \dot{d}_{k\alpha}^\dag c_i + {\rm H.c.} \right>} 
  = -\overline{\left< I_{\alpha B} \right>} .
\end{equation}
The above equality is only valid when taking the time average since the couplings between
the subsystems and the electronic reservoirs periodically store and release energy in
response to the driving field~\cite{Sanchez14}.
For DC components, we thus obtain 
\begin{equation}
  \bar{I}_{\alpha C}
  = -\bar{I}_{\alpha B} +\bar{I}_{\alpha} -\bar{I}_{\alpha D} ,
\end{equation}
where we have used 
\begin{equation}
  [\hat{H}_{\phi}, \hat{H}_{\alpha C}] 
  = \big[\hat{H}_{\phi}, \hat{V}_{\alpha} \big] 
  = -e_0 \sum_{i\in\alpha} \rho_i \big[\hat{H}_{\rm tot}, \phi(\bm{r}_i) \big] .
\end{equation}
Since $\hat{H}_{\alpha C}$ has finite degrees of freedom, we have $\bar{I}_{\alpha C}=0$
using Eq.~\eqref{avg}. Thus, the energy conservation described in Eq.~\eqref{conservation}
is proved.

\section{Proof of energy conservation using NEGF}
\renewcommand{\theequation}{E.\arabic{equation}}
We prove the energy conservation $\bar{I}_{\alpha} =\bar{I}_{\alpha B}+\bar{I}_{\alpha D}$
in the framework of NEGF. Equation~\eqref{Dyson_G} can be written as
\begin{equation}
  G = g + G (\Sigma_B + \Sigma_{\phi} + \Sigma_H) g ,
\end{equation}
where $g$ is the Green's function of the isolated subsystem. 
The corresponding lesser component is
\begin{equation}
  G^< g^{-1} = \big[G (\Sigma_B + \Sigma_{\phi} + \Sigma_H)\big]^< ,
\end{equation}
which can be transformed to the energy space as
\begin{equation}
  \bm{G}^< (\bm{E}-\bm{h}) = \big[\bm{G} (\bm{\Sigma}_B + \bm{\Sigma}_{\phi}
  + \bm{\Sigma}_H)\big]^< .
\end{equation}
Using this relation, we add Eq.~\eqref{IB} and Eq.~\eqref{ID} to obtain
\begin{align}
  & \bar{I}_{\alpha B}+\bar{I}_{\alpha D}  \notag \\
  =& -\int_{\rm BZ} \frac{dE}{2\pi\hbar} {\rm Tr}_{\alpha} \Big[ \bm{E}
  (\bm{G}\bm{\Sigma}_{\phi}+ \bm{G}\bm{\Sigma}_H)^< + {\rm H.c.} \Big]
  \notag \\
  =& -\int_{\rm BZ} \frac{dE}{2\pi\hbar} {\rm Tr}_{\alpha}\Big[ \bm{E} (\bm{G}^r
  \bm{\Sigma}_{\phi}^< + \bm{G}^< \bm{\Sigma}_{\phi}^a) + {\rm H.c.} \Big] ,
\end{align}
where ${\rm Tr}_{\alpha}$ denotes trace over electronic sites in subsystem $\alpha$ and
Floquet space. The Hartree self-energy does not contribute to energy transport because of
the symmetry of $\bm{G}^<$ shown in Eq.~\eqref{sym_GE}, and the fact that
$\bm{\Sigma}_H^a$ is a real diagonal matrix. Switching to the time domain, we obtain
\begin{align} \label{IBID}
  & \bar{I}_{\alpha B}+\bar{I}_{\alpha D} = -\frac{i\hbar}{T} \int_0^T dt_1
  \int_{-\infty}^{\infty} dt_2 \notag \\
  & {\rm tr}_{\alpha} \Big[ 
  \partial_{t_1} G^r(t_1,t_2)\Sigma_{\phi}^<(t_2,t_1) +
  \partial_{t_1} G^<(t_1,t_2)\Sigma_{\phi}^a(t_2,t_1)
  \notag \\
  &\ \ + \Sigma_{\phi}^<(t_1,t_2) \partial_{t_1} G^a(t_2,t_1)
  + \Sigma_{\phi}^r(t_1,t_2) \partial_{t_1} G^<(t_2,t_1) \Big] ,
\end{align}
where ${\rm tr}_{\alpha}$ traces over electronic sites in part $\alpha$.
We can write the terms in ${\rm tr}_{\alpha}[\cdots ]$ as
\begin{align} \label{GGW}
  i\hbar \Big[
   & \partial_{t_1} G^r(t_1,t_2) G^<(t_2,t_1) W^<(t_2,t_1) \notag \\
  +& \partial_{t_1} G^<(t_1,t_2) G^a(t_2,t_1) W^<(t_2,t_1) \notag \\
  +& \partial_{t_1} G^<(t_1,t_2) G^>(t_2,t_1) W^a(t_2,t_1) \notag \\
  +& W^>(t_2,t_1) G^<(t_1,t_2) \partial_{t_1} G^a(t_2,t_1) \notag \\
  +& W^>(t_2,t_1) G^r(t_1,t_2) \partial_{t_1} G^<(t_2,t_1) \notag \\
  +& W^a(t_2,t_1) G^>(t_1,t_2) \partial_{t_1} G^<(t_2,t_1) \Big] .
\end{align}
The fourth to sixth terms above are obtained by taking Hermitian conjugate of the first to
third terms and using the symmetry shown in Eq.~\eqref{sym_W} and Eq.~\eqref{sym_G} for
$W$. By omitting the factor $i\hbar$, we can symbolically write above terms as
\begin{align} \label{symbolic}
  &  (\partial_1 G^r_{12}) G^<_{21} W^<_{21} 
  +  (\partial_1 G^<_{12}) G^a_{21} W^<_{21} 
  +  (\partial_1 G^<_{12}) G^>_{21} W^a_{21}  \notag \\
  +& G^<_{12} (\partial_1 G^a_{21}) W^>_{21}
  +  G^r_{12} (\partial_1 G^<_{21}) W^>_{21} 
  +  G^>_{12} (\partial_1 G^<_{21}) W^a_{21} ,
\end{align}
where the subscripts $1$ and $2$ represent time variables $t_1$ and $t_2$ respectively. 
Taking the sum of the first and fifth terms of Eq.~\eqref{symbolic} yields
\begin{align} \label{15}
  & (\partial_1 G^r_{12}) G^<_{21} W^<_{21} + G^r_{12} (\partial_1 G^<_{21}) W^>_{21} 
  \notag \\
  =& [(\partial_1 G^r_{12}) G^<_{21} + G^r_{12} (\partial_1 G^<_{21})] W^<_{21} 
  - G^r_{12} (\partial_1 G^<_{21}) W^a_{21} .
\end{align}
where the term $G^r_{12} (\partial_1 G^<_{21}) W^r_{21}$  is omitted as it vanishes.
Adding the second and the fourth terms of Eq.~\eqref{symbolic} results in
\begin{align}
  &(\partial_1 G^<_{12}) G^a_{21} W^<_{21} + G^<_{12} (\partial_1 G^a_{21}) W^>_{21} 
  \notag \\
  =& [(\partial_1 G^<_{12}) G^a_{21} + G^<_{12} (\partial_1 G^a_{21})] W^<_{21} 
  - G^<_{12} (\partial_1 G^a_{21}) W^a_{21} \notag \\
  &+ G^<_{12} (\partial_1 G^a_{21}) W^r_{21} .
\end{align}
Adding the first, the second, the fourth, and the fifth terms of Eq.~\eqref{GGW} thus
yields
\begin{align} \label{1245}
  &- (\partial_1 \Pi^r_{12}) W^<_{21} 
  - i\hbar [G^r_{12}(\partial_1 G^<_{21}) +G^<_{12}(\partial_1 G^a_{21})] W^a_{21}
  \notag \\
  &+ i\hbar G^<_{12} (\partial_1 G^a_{21}) W^r_{21} .
\end{align}
Similarly to the procedure above, by adding the third and the sixth terms of
Eq.~\eqref{GGW}, we obtain
\begin{align} \label{36}
  &- (\partial_1 \Pi^<_{12}) W^a_{21} 
  + i\hbar [G^r_{12}(\partial_1 G^<_{21}) + G^<_{12}(\partial_1 G^a_{21})] W^a_{21}
  \notag \\
  &- i\hbar G^<_{12} (\partial_1 G^r_{21}) W^a_{21} ,
\end{align}
where we have used
\begin{align}
  G^>_{12} (\partial_1 G^<_{21}) = G^<_{12} \partial_1 G^>_{21} 
  + & G^<_{12} (\partial_1 G^a_{21} -\partial_1 G^r_{21}) \notag \\
  + & (G^r_{12} - G^a_{12}) \partial_1 G^<_{21} ,
\end{align}
and omitted the vanishing term $-G^a_{12}(\partial_1 G^<_{21}) W^a_{21}$. 
By taking the sum of Eqs.~\eqref{1245} and \eqref{36}, Eq.~\eqref{GGW} is equivalent to
\begin{align} \label{123456}
  &- (\partial_1 \Pi^r_{12}) W^<_{21} - (\partial_1 \Pi^<_{12}) W^a_{21} \notag \\
  &+ i\hbar G^<_{12} \big[(\partial_1 G^a_{21}) W^r_{21} 
  - (\partial_1 G^r_{21}) W^a_{21} \big] .
\end{align}
Using the fact that
\begin{equation}
  G^a_{21} = -\theta(t_1-t_2) (G^>_{21} - G^<_{21}) , \
  G^r_{21} = \theta(t_2-t_1) (G^>_{21} - G^<_{21}) ,
\end{equation}
one has
\begin{align} \label{GGWW}
  & (\partial_1 G^a_{21}) W^r_{21} - (\partial_1 G^r_{21}) W^a_{21} \notag \\
  =& -\delta(t_1-t_2) (G^>_{21} - G^<_{21}) (W^r_{21} - W^a_{21}) .
\end{align}
Therefore, we arrive at 
\begin{align} \label{IBID_1}
  & \bar{I}_{\alpha B}+\bar{I}_{\alpha D} = I_i + \frac{i\hbar}{T} \int_0^T dt_1
  \int_{-\infty}^{\infty} dt_2 \notag \\
  & {\rm tr}_{\alpha} \Big[ \partial_{t_1} \Pi^r(t_1,t_2) W^<(t_2,t_1) +
  \partial_{t_1} \Pi^<(t_1,t_2) W^a(t_2,t_1) \Big] ,
\end{align}
where the purely imaginary term $I_i$ comes from the second line in Eq.~\eqref{123456}. 
Using the symmetry in Eq.~\eqref{sym_W}, Eq.~\eqref{symbolic} is equivalent to
\begin{align} \label{symbolic2}
  &  W^>_{12} (\partial_1 G^r_{12}) G^<_{21} 
  +  W^>_{12} (\partial_1 G^<_{12}) G^a_{21} 
  +  W^r_{12} (\partial_1 G^<_{12}) G^>_{21}  \notag \\
  +& W^<_{12} G^<_{12} (\partial_1 G^a_{21}) 
  +  W^<_{12} G^r_{12} (\partial_1 G^<_{21}) 
  +  W^r_{12} G^>_{12} (\partial_1 G^<_{21}) .
\end{align}
Similarly to the steps from Eq.~\eqref{15} to Eq.~\eqref{GGWW}, we have
\begin{align} \label{IBID_2}
  & \bar{I}_{\alpha B}+\bar{I}_{\alpha D} = -I_i + \frac{i\hbar}{T} \int_0^T dt_1
  \int_{-\infty}^{\infty} dt_2 \notag \\
  & {\rm tr}_{\alpha} \Big[  W^<(t_1,t_2) \partial_{t_1} \Pi^a(t_2,t_1) +
  \partial_{t_1} \Pi^<(t_2,t_1) W^r(t_1,t_2) \Big] .
\end{align}
By adding Eq.~\eqref{IBID_1} and \eqref{IBID_2}, we eliminate the term $I_i$ and arrive at
\begin{align}
  \bar{I}_{\alpha B} & +\bar{I}_{\alpha D} = \frac{1}{2} \frac{i\hbar}{T} \int_0^T dt_1
  \int_{-\infty}^{\infty} dt_2 \notag \\
  {\rm tr}_{\alpha} \Big[ & \partial_{t_1} \Pi^r(t_1,t_2) W^<(t_2,t_1) +
  \partial_{t_1} \Pi^<(t_1,t_2) W^a(t_2,t_1) \notag \\
  +& W^<(t_1,t_2) \partial_{t_1} \Pi^a(t_2,t_1) +
  W^r(t_1,t_2) \partial_{t_1} \Pi^<(t_2,t_1) \Big] ,
\end{align}
which can be expressed in the energy domain as
\begin{align} 
  \bar{I}_{\alpha B}+\bar{I}_{\alpha D} = \frac{1}{2} \int_{\rm BZ} \frac{dE}{2\pi\hbar}
  {\rm Tr}_{\alpha} \big[\bm{E} \big( 
    & \bm{\Pi}^r \bm{W}^< + \bm{\Pi}^< \bm{W}^a  \notag \\
  - & \bm{W}^< \bm{\Pi}^a - \bm{W}^r \bm{\Pi}^< \big)\big] .
\end{align}
Thus, the energy conservation regarding the DC currents $\bar{I}_{\alpha} =
\bar{I}_{\alpha B} + \bar{I}_{\alpha D}$ is proved from the perspective of NEGF. 
It can be demonstrated that this conservation is still upheld under the $G_0W_0$
approximation.

\bibliography{bib_heat_radiation}{}

\begin{thebibliography}{53}%
\makeatletter
\providecommand \@ifxundefined [1]{%
 \@ifx{#1\undefined}
}%
\providecommand \@ifnum [1]{%
 \ifnum #1\expandafter \@firstoftwo
 \else \expandafter \@secondoftwo
 \fi
}%
\providecommand \@ifx [1]{%
 \ifx #1\expandafter \@firstoftwo
 \else \expandafter \@secondoftwo
 \fi
}%
\providecommand \natexlab [1]{#1}%
\providecommand \enquote  [1]{``#1''}%
\providecommand \bibnamefont  [1]{#1}%
\providecommand \bibfnamefont [1]{#1}%
\providecommand \citenamefont [1]{#1}%
\providecommand \href@noop [0]{\@secondoftwo}%
\providecommand \href [0]{\begingroup \@sanitize@url \@href}%
\providecommand \@href[1]{\@@startlink{#1}\@@href}%
\providecommand \@@href[1]{\endgroup#1\@@endlink}%
\providecommand \@sanitize@url [0]{\catcode `\\12\catcode `\$12\catcode
  `\&12\catcode `\#12\catcode `\^12\catcode `\_12\catcode `\%12\relax}%
\providecommand \@@startlink[1]{}%
\providecommand \@@endlink[0]{}%
\providecommand \url  [0]{\begingroup\@sanitize@url \@url }%
\providecommand \@url [1]{\endgroup\@href {#1}{\urlprefix }}%
\providecommand \urlprefix  [0]{URL }%
\providecommand \Eprint [0]{\href }%
\providecommand \doibase [0]{https://doi.org/}%
\providecommand \selectlanguage [0]{\@gobble}%
\providecommand \bibinfo  [0]{\@secondoftwo}%
\providecommand \bibfield  [0]{\@secondoftwo}%
\providecommand \translation [1]{[#1]}%
\providecommand \BibitemOpen [0]{}%
\providecommand \bibitemStop [0]{}%
\providecommand \bibitemNoStop [0]{.\EOS\space}%
\providecommand \EOS [0]{\spacefactor3000\relax}%
\providecommand \BibitemShut  [1]{\csname bibitem#1\endcsname}%
\let\auto@bib@innerbib\@empty
\bibitem [{\citenamefont {Pendry}(1999)}]{Pendry99}%
  \BibitemOpen
  \bibfield  {author} {\bibinfo {author} {\bibfnamefont {J.~B.}\ \bibnamefont
  {Pendry}},\ }\bibfield  {title} {\bibinfo {title} {Radiative exchange of heat
  between nanostructures},\ }\href
  {https://doi.org/10.1088/0953-8984/11/35/301} {\bibfield  {journal} {\bibinfo
   {journal} {J. Phys.: Condens. Matter}\ }\textbf {\bibinfo {volume} {11}},\
  \bibinfo {pages} {6621} (\bibinfo {year} {1999})}\BibitemShut {NoStop}%
\bibitem [{\citenamefont {Volokitin}\ and\ \citenamefont
  {Persson}(2007)}]{review07}%
  \BibitemOpen
  \bibfield  {author} {\bibinfo {author} {\bibfnamefont {A.~I.}\ \bibnamefont
  {Volokitin}}\ and\ \bibinfo {author} {\bibfnamefont {B.~N.~J.}\ \bibnamefont
  {Persson}},\ }\bibfield  {title} {\bibinfo {title} {Near-field radiative heat
  transfer and noncontact friction},\ }\href
  {https://doi.org/10.1103/RevModPhys.79.1291} {\bibfield  {journal} {\bibinfo
  {journal} {Rev. Mod. Phys.}\ }\textbf {\bibinfo {volume} {79}},\ \bibinfo
  {pages} {1291} (\bibinfo {year} {2007})}\BibitemShut {NoStop}%
\bibitem [{\citenamefont {Song}\ \emph {et~al.}(2015)\citenamefont {Song},
  \citenamefont {Fiorino}, \citenamefont {Meyhofer},\ and\ \citenamefont
  {Reddy}}]{review15}%
  \BibitemOpen
  \bibfield  {author} {\bibinfo {author} {\bibfnamefont {B.}~\bibnamefont
  {Song}}, \bibinfo {author} {\bibfnamefont {A.}~\bibnamefont {Fiorino}},
  \bibinfo {author} {\bibfnamefont {E.}~\bibnamefont {Meyhofer}},\ and\
  \bibinfo {author} {\bibfnamefont {P.}~\bibnamefont {Reddy}},\ }\bibfield
  {title} {\bibinfo {title} {Near-field radiative thermal transport: From
  theory to experiment},\ }\href {https://doi.org/10.1063/1.4919048} {\bibfield
   {journal} {\bibinfo  {journal} {AIP Adv.}\ }\textbf {\bibinfo {volume}
  {5}},\ \bibinfo {pages} {053503} (\bibinfo {year} {2015})}\BibitemShut
  {NoStop}%
\bibitem [{\citenamefont {Liu}\ \emph {et~al.}(2015)\citenamefont {Liu},
  \citenamefont {Wang},\ and\ \citenamefont {Zhang}}]{review15-2}%
  \BibitemOpen
  \bibfield  {author} {\bibinfo {author} {\bibfnamefont {X.}~\bibnamefont
  {Liu}}, \bibinfo {author} {\bibfnamefont {L.}~\bibnamefont {Wang}},\ and\
  \bibinfo {author} {\bibfnamefont {Z.~M.}\ \bibnamefont {Zhang}},\ }\bibfield
  {title} {\bibinfo {title} {Near-field thermal radiation: Recent progress and
  outlook},\ }\href {https://doi.org/10.1080/15567265.2015.1027836} {\bibfield
  {journal} {\bibinfo  {journal} {Nanoscale and Microscale Thermophys. Eng.}\
  }\textbf {\bibinfo {volume} {19}},\ \bibinfo {pages} {98} (\bibinfo {year}
  {2015})}\BibitemShut {NoStop}%
\bibitem [{\citenamefont {Cuevas}\ and\ \citenamefont
  {García-Vidal}(2018)}]{review18}%
  \BibitemOpen
  \bibfield  {author} {\bibinfo {author} {\bibfnamefont {J.~C.}\ \bibnamefont
  {Cuevas}}\ and\ \bibinfo {author} {\bibfnamefont {F.~J.}\ \bibnamefont
  {García-Vidal}},\ }\bibfield  {title} {\bibinfo {title} {Radiative heat
  transfer},\ }\href {https://doi.org/10.1021/acsphotonics.8b01031} {\bibfield
  {journal} {\bibinfo  {journal} {ACS Photonics}\ }\textbf {\bibinfo {volume}
  {5}},\ \bibinfo {pages} {3896} (\bibinfo {year} {2018})}\BibitemShut
  {NoStop}%
\bibitem [{\citenamefont {Biehs}\ \emph {et~al.}(2021)\citenamefont {Biehs},
  \citenamefont {Messina}, \citenamefont {Venkataram}, \citenamefont
  {Rodriguez}, \citenamefont {Cuevas},\ and\ \citenamefont
  {Ben-Abdallah}}]{review20}%
  \BibitemOpen
  \bibfield  {author} {\bibinfo {author} {\bibfnamefont {S.-A.}\ \bibnamefont
  {Biehs}}, \bibinfo {author} {\bibfnamefont {R.}~\bibnamefont {Messina}},
  \bibinfo {author} {\bibfnamefont {P.~S.}\ \bibnamefont {Venkataram}},
  \bibinfo {author} {\bibfnamefont {A.~W.}\ \bibnamefont {Rodriguez}}, \bibinfo
  {author} {\bibfnamefont {J.~C.}\ \bibnamefont {Cuevas}},\ and\ \bibinfo
  {author} {\bibfnamefont {P.}~\bibnamefont {Ben-Abdallah}},\ }\bibfield
  {title} {\bibinfo {title} {Near-field radiative heat transfer in many-body
  systems},\ }\href {https://doi.org/10.1103/RevModPhys.93.025009} {\bibfield
  {journal} {\bibinfo  {journal} {Rev. Mod. Phys.}\ }\textbf {\bibinfo {volume}
  {93}},\ \bibinfo {pages} {025009} (\bibinfo {year} {2021})}\BibitemShut
  {NoStop}%
\bibitem [{\citenamefont {Tang}\ \emph {et~al.}(2021)\citenamefont {Tang},
  \citenamefont {Zhang}, \citenamefont {Zhang}, \citenamefont {Chen},\ and\
  \citenamefont {Chan}}]{GT21}%
  \BibitemOpen
  \bibfield  {author} {\bibinfo {author} {\bibfnamefont {G.}~\bibnamefont
  {Tang}}, \bibinfo {author} {\bibfnamefont {L.}~\bibnamefont {Zhang}},
  \bibinfo {author} {\bibfnamefont {Y.}~\bibnamefont {Zhang}}, \bibinfo
  {author} {\bibfnamefont {J.}~\bibnamefont {Chen}},\ and\ \bibinfo {author}
  {\bibfnamefont {C.~T.}\ \bibnamefont {Chan}},\ }\bibfield  {title} {\bibinfo
  {title} {Near-field energy transfer between graphene and magneto-optic
  media},\ }\href {https://doi.org/10.1103/PhysRevLett.127.247401} {\bibfield
  {journal} {\bibinfo  {journal} {Phys. Rev. Lett.}\ }\textbf {\bibinfo
  {volume} {127}},\ \bibinfo {pages} {247401} (\bibinfo {year}
  {2021})}\BibitemShut {NoStop}%
\bibitem [{\citenamefont {Rytov}(1953)}]{Rytov}%
  \BibitemOpen
  \bibfield  {author} {\bibinfo {author} {\bibfnamefont {S.~M.}\ \bibnamefont
  {Rytov}},\ }\href@noop {} {\emph {\bibinfo {title} {Theory of Electrical
  Fluctuation and Thermal Radiation}}}\ (\bibinfo  {publisher} {Academy of
  Science of USSR, Moscow},\ \bibinfo {year} {1953})\BibitemShut {NoStop}%
\bibitem [{\citenamefont {Polder}\ and\ \citenamefont {Van~Hove}(1971)}]{PvH}%
  \BibitemOpen
  \bibfield  {author} {\bibinfo {author} {\bibfnamefont {D.}~\bibnamefont
  {Polder}}\ and\ \bibinfo {author} {\bibfnamefont {M.}~\bibnamefont
  {Van~Hove}},\ }\bibfield  {title} {\bibinfo {title} {Theory of radiative heat
  transfer between closely spaced bodies},\ }\href
  {https://doi.org/10.1103/PhysRevB.4.3303} {\bibfield  {journal} {\bibinfo
  {journal} {Phys. Rev. B}\ }\textbf {\bibinfo {volume} {4}},\ \bibinfo {pages}
  {3303} (\bibinfo {year} {1971})}\BibitemShut {NoStop}%
\bibitem [{\citenamefont {Coppens}\ and\ \citenamefont
  {Valentine}(2017)}]{coppens17}%
  \BibitemOpen
  \bibfield  {author} {\bibinfo {author} {\bibfnamefont {Z.~J.}\ \bibnamefont
  {Coppens}}\ and\ \bibinfo {author} {\bibfnamefont {J.~G.}\ \bibnamefont
  {Valentine}},\ }\bibfield  {title} {\bibinfo {title} {Spatial and temporal
  modulation of thermal emission},\ }\href@noop {} {\bibfield  {journal}
  {\bibinfo  {journal} {Adv. Mater.}\ }\textbf {\bibinfo {volume} {29}},\
  \bibinfo {pages} {1701275} (\bibinfo {year} {2017})}\BibitemShut {NoStop}%
\bibitem [{\citenamefont {Latella}\ \emph {et~al.}(2018)\citenamefont
  {Latella}, \citenamefont {Messina}, \citenamefont {Rubi},\ and\ \citenamefont
  {Ben-Abdallah}}]{Shuttling18}%
  \BibitemOpen
  \bibfield  {author} {\bibinfo {author} {\bibfnamefont {I.}~\bibnamefont
  {Latella}}, \bibinfo {author} {\bibfnamefont {R.}~\bibnamefont {Messina}},
  \bibinfo {author} {\bibfnamefont {J.~M.}\ \bibnamefont {Rubi}},\ and\
  \bibinfo {author} {\bibfnamefont {P.}~\bibnamefont {Ben-Abdallah}},\
  }\bibfield  {title} {\bibinfo {title} {Radiative heat shuttling},\ }\href
  {https://doi.org/10.1103/PhysRevLett.121.023903} {\bibfield  {journal}
  {\bibinfo  {journal} {Phys. Rev. Lett.}\ }\textbf {\bibinfo {volume} {121}},\
  \bibinfo {pages} {023903} (\bibinfo {year} {2018})}\BibitemShut {NoStop}%
\bibitem [{\citenamefont {Kou}\ and\ \citenamefont {Minnich}(2018)}]{Kou18}%
  \BibitemOpen
  \bibfield  {author} {\bibinfo {author} {\bibfnamefont {J.}~\bibnamefont
  {Kou}}\ and\ \bibinfo {author} {\bibfnamefont {A.~J.}\ \bibnamefont
  {Minnich}},\ }\bibfield  {title} {\bibinfo {title} {Dynamic optical control
  of near-field radiative transfer},\ }\href
  {https://doi.org/10.1364/OE.26.00A729} {\bibfield  {journal} {\bibinfo
  {journal} {Opt. Express}\ }\textbf {\bibinfo {volume} {26}},\ \bibinfo
  {pages} {A729} (\bibinfo {year} {2018})}\BibitemShut {NoStop}%
\bibitem [{\citenamefont {Li}\ \emph {et~al.}(2019)\citenamefont {Li},
  \citenamefont {Fern\'andez-Alc\'azar}, \citenamefont {Ellis}, \citenamefont
  {Shapiro},\ and\ \citenamefont {Kottos}}]{Li19}%
  \BibitemOpen
  \bibfield  {author} {\bibinfo {author} {\bibfnamefont {H.}~\bibnamefont
  {Li}}, \bibinfo {author} {\bibfnamefont {L.~J.}\ \bibnamefont
  {Fern\'andez-Alc\'azar}}, \bibinfo {author} {\bibfnamefont {F.}~\bibnamefont
  {Ellis}}, \bibinfo {author} {\bibfnamefont {B.}~\bibnamefont {Shapiro}},\
  and\ \bibinfo {author} {\bibfnamefont {T.}~\bibnamefont {Kottos}},\
  }\bibfield  {title} {\bibinfo {title} {Adiabatic thermal radiation pumps for
  thermal photonics},\ }\href {https://doi.org/10.1103/PhysRevLett.123.165901}
  {\bibfield  {journal} {\bibinfo  {journal} {Phys. Rev. Lett.}\ }\textbf
  {\bibinfo {volume} {123}},\ \bibinfo {pages} {165901} (\bibinfo {year}
  {2019})}\BibitemShut {NoStop}%
\bibitem [{\citenamefont {Buddhiraju}\ \emph {et~al.}(2020)\citenamefont
  {Buddhiraju}, \citenamefont {Li},\ and\ \citenamefont {Fan}}]{Fan20}%
  \BibitemOpen
  \bibfield  {author} {\bibinfo {author} {\bibfnamefont {S.}~\bibnamefont
  {Buddhiraju}}, \bibinfo {author} {\bibfnamefont {W.}~\bibnamefont {Li}},\
  and\ \bibinfo {author} {\bibfnamefont {S.}~\bibnamefont {Fan}},\ }\bibfield
  {title} {\bibinfo {title} {Photonic refrigeration from time-modulated thermal
  emission},\ }\href {https://doi.org/10.1103/PhysRevLett.124.077402}
  {\bibfield  {journal} {\bibinfo  {journal} {Phys. Rev. Lett.}\ }\textbf
  {\bibinfo {volume} {124}},\ \bibinfo {pages} {077402} (\bibinfo {year}
  {2020})}\BibitemShut {NoStop}%
\bibitem [{\citenamefont {Fern\'andez-Alc\'azar}\ \emph
  {et~al.}(2021{\natexlab{a}})\citenamefont {Fern\'andez-Alc\'azar},
  \citenamefont {Kononchuk}, \citenamefont {Li},\ and\ \citenamefont
  {Kottos}}]{Li21}%
  \BibitemOpen
  \bibfield  {author} {\bibinfo {author} {\bibfnamefont {L.~J.}\ \bibnamefont
  {Fern\'andez-Alc\'azar}}, \bibinfo {author} {\bibfnamefont {R.}~\bibnamefont
  {Kononchuk}}, \bibinfo {author} {\bibfnamefont {H.}~\bibnamefont {Li}},\ and\
  \bibinfo {author} {\bibfnamefont {T.}~\bibnamefont {Kottos}},\ }\bibfield
  {title} {\bibinfo {title} {Extreme nonreciprocal near-field thermal radiation
  via {F}loquet photonics},\ }\href
  {https://doi.org/10.1103/PhysRevLett.126.204101} {\bibfield  {journal}
  {\bibinfo  {journal} {Phys. Rev. Lett.}\ }\textbf {\bibinfo {volume} {126}},\
  \bibinfo {pages} {204101} (\bibinfo {year} {2021}{\natexlab{a}})}\BibitemShut
  {NoStop}%
\bibitem [{\citenamefont {Fern\'andez-Alc\'azar}\ \emph
  {et~al.}(2021{\natexlab{b}})\citenamefont {Fern\'andez-Alc\'azar},
  \citenamefont {Li}, \citenamefont {Nafari},\ and\ \citenamefont
  {Kottos}}]{Li21_ACS}%
  \BibitemOpen
  \bibfield  {author} {\bibinfo {author} {\bibfnamefont {L.~J.}\ \bibnamefont
  {Fern\'andez-Alc\'azar}}, \bibinfo {author} {\bibfnamefont {H.}~\bibnamefont
  {Li}}, \bibinfo {author} {\bibfnamefont {M.}~\bibnamefont {Nafari}},\ and\
  \bibinfo {author} {\bibfnamefont {T.}~\bibnamefont {Kottos}},\ }\bibfield
  {title} {\bibinfo {title} {Implementation of optimal thermal radiation pumps
  using adiabatically modulated photonic cavities},\ }\href
  {https://doi.org/10.1021/acsphotonics.1c00896} {\bibfield  {journal}
  {\bibinfo  {journal} {ACS Photonics}\ }\textbf {\bibinfo {volume} {8}},\
  \bibinfo {pages} {2973} (\bibinfo {year} {2021}{\natexlab{b}})}\BibitemShut
  {NoStop}%
\bibitem [{\citenamefont {Yu}\ and\ \citenamefont
  {Fan}(2023{\natexlab{a}})}]{Fan23_1}%
  \BibitemOpen
  \bibfield  {author} {\bibinfo {author} {\bibfnamefont {R.}~\bibnamefont
  {Yu}}\ and\ \bibinfo {author} {\bibfnamefont {S.}~\bibnamefont {Fan}},\
  }\bibfield  {title} {\bibinfo {title} {Manipulating coherence of near-field
  thermal radiation in time-modulated systems},\ }\href
  {https://doi.org/10.1103/PhysRevLett.130.096902} {\bibfield  {journal}
  {\bibinfo  {journal} {Phys. Rev. Lett.}\ }\textbf {\bibinfo {volume} {130}},\
  \bibinfo {pages} {096902} (\bibinfo {year} {2023}{\natexlab{a}})}\BibitemShut
  {NoStop}%
\bibitem [{\citenamefont {Yu}\ and\ \citenamefont
  {Fan}(2023{\natexlab{b}})}]{Fan23_2}%
  \BibitemOpen
  \bibfield  {author} {\bibinfo {author} {\bibfnamefont {R.}~\bibnamefont
  {Yu}}\ and\ \bibinfo {author} {\bibfnamefont {S.}~\bibnamefont {Fan}},\
  }\href@noop {} {\bibinfo {title} {Time-modulated near-field radiative heat
  transfer}} (\bibinfo {year} {2023}{\natexlab{b}}),\ \Eprint
  {https://arxiv.org/abs/arXiv:2310.08692} {arXiv:2310.08692} \BibitemShut
  {NoStop}%
\bibitem [{\citenamefont {Biehs}\ and\ \citenamefont
  {Agarwal}(2023{\natexlab{a}})}]{Biehs22}%
  \BibitemOpen
  \bibfield  {author} {\bibinfo {author} {\bibfnamefont {S.-A.}\ \bibnamefont
  {Biehs}}\ and\ \bibinfo {author} {\bibfnamefont {G.~S.}\ \bibnamefont
  {Agarwal}},\ }\bibfield  {title} {\bibinfo {title} {Breakdown of detailed
  balance for thermal radiation by synthetic fields},\ }\href
  {https://doi.org/10.1103/PhysRevLett.130.110401} {\bibfield  {journal}
  {\bibinfo  {journal} {Phys. Rev. Lett.}\ }\textbf {\bibinfo {volume} {130}},\
  \bibinfo {pages} {110401} (\bibinfo {year} {2023}{\natexlab{a}})}\BibitemShut
  {NoStop}%
\bibitem [{\citenamefont {Biehs}\ and\ \citenamefont
  {Agarwal}(2023{\natexlab{b}})}]{Biehs23_1}%
  \BibitemOpen
  \bibfield  {author} {\bibinfo {author} {\bibfnamefont {S.-A.}\ \bibnamefont
  {Biehs}}\ and\ \bibinfo {author} {\bibfnamefont {G.~S.}\ \bibnamefont
  {Agarwal}},\ }\bibfield  {title} {\bibinfo {title} {Enhancement of synthetic
  magnetic field induced nonreciprocity via bound states in the continuum in
  dissipatively coupled systems},\ }\href
  {https://doi.org/10.1103/PhysRevB.108.035423} {\bibfield  {journal} {\bibinfo
   {journal} {Phys. Rev. B}\ }\textbf {\bibinfo {volume} {108}},\ \bibinfo
  {pages} {035423} (\bibinfo {year} {2023}{\natexlab{b}})}\BibitemShut
  {NoStop}%
\bibitem [{\citenamefont {Biehs}\ \emph {et~al.}(2023)\citenamefont {Biehs},
  \citenamefont {Rodriguez-Lopez}, \citenamefont {Antezza},\ and\ \citenamefont
  {Agarwal}}]{Biehs23_2}%
  \BibitemOpen
  \bibfield  {author} {\bibinfo {author} {\bibfnamefont {S.-A.}\ \bibnamefont
  {Biehs}}, \bibinfo {author} {\bibfnamefont {P.}~\bibnamefont
  {Rodriguez-Lopez}}, \bibinfo {author} {\bibfnamefont {M.}~\bibnamefont
  {Antezza}},\ and\ \bibinfo {author} {\bibfnamefont {G.~S.}\ \bibnamefont
  {Agarwal}},\ }\bibfield  {title} {\bibinfo {title} {Nonreciprocal heat flux
  via synthetic fields in linear quantum systems},\ }\href
  {https://doi.org/10.1103/PhysRevA.108.042201} {\bibfield  {journal} {\bibinfo
   {journal} {Phys. Rev. A}\ }\textbf {\bibinfo {volume} {108}},\ \bibinfo
  {pages} {042201} (\bibinfo {year} {2023})}\BibitemShut {NoStop}%
\bibitem [{\citenamefont {V{\'a}zquez-Lozano}\ and\ \citenamefont
  {Liberal}(2023)}]{Lozano23}%
  \BibitemOpen
  \bibfield  {author} {\bibinfo {author} {\bibfnamefont {J.~E.}\ \bibnamefont
  {V{\'a}zquez-Lozano}}\ and\ \bibinfo {author} {\bibfnamefont
  {I.}~\bibnamefont {Liberal}},\ }\bibfield  {title} {\bibinfo {title}
  {Incandescent temporal metamaterials},\ }\href
  {https://doi.org/10.1038/s41467-023-40281-2} {\bibfield  {journal} {\bibinfo
  {journal} {Nat. Commun.}\ }\textbf {\bibinfo {volume} {14}},\ \bibinfo
  {pages} {4606} (\bibinfo {year} {2023})}\BibitemShut {NoStop}%
\bibitem [{\citenamefont {Picardi}\ \emph {et~al.}(2023)\citenamefont
  {Picardi}, \citenamefont {Nimje},\ and\ \citenamefont
  {Papadakis}}]{Picardi23}%
  \BibitemOpen
  \bibfield  {author} {\bibinfo {author} {\bibfnamefont {M.~F.}\ \bibnamefont
  {Picardi}}, \bibinfo {author} {\bibfnamefont {K.~N.}\ \bibnamefont {Nimje}},\
  and\ \bibinfo {author} {\bibfnamefont {G.~T.}\ \bibnamefont {Papadakis}},\
  }\bibfield  {title} {\bibinfo {title} {Dynamic modulation of thermal
  emission—{A} tutorial},\ }\href {https://doi.org/10.1063/5.0134951}
  {\bibfield  {journal} {\bibinfo  {journal} {J. Appl. Phys.}\ }\textbf
  {\bibinfo {volume} {133}},\ \bibinfo {pages} {111101} (\bibinfo {year}
  {2023})}\BibitemShut {NoStop}%
\bibitem [{\citenamefont {Wang}\ and\ \citenamefont {Peng}(2017)}]{JSW0}%
  \BibitemOpen
  \bibfield  {author} {\bibinfo {author} {\bibfnamefont {J.-S.}\ \bibnamefont
  {Wang}}\ and\ \bibinfo {author} {\bibfnamefont {J.}~\bibnamefont {Peng}},\
  }\bibfield  {title} {\bibinfo {title} {Capacitor physics in ultra-near-field
  heat transfer},\ }\href {https://doi.org/10.1209/0295-5075/118/24001}
  {\bibfield  {journal} {\bibinfo  {journal} {{EPL} (Europhysics Letters)}\
  }\textbf {\bibinfo {volume} {118}},\ \bibinfo {pages} {24001} (\bibinfo
  {year} {2017})}\BibitemShut {NoStop}%
\bibitem [{\citenamefont {Jiang}\ and\ \citenamefont {Wang}(2017)}]{JSW1}%
  \BibitemOpen
  \bibfield  {author} {\bibinfo {author} {\bibfnamefont {J.-H.}\ \bibnamefont
  {Jiang}}\ and\ \bibinfo {author} {\bibfnamefont {J.-S.}\ \bibnamefont
  {Wang}},\ }\bibfield  {title} {\bibinfo {title} {Caroli formalism in
  near-field heat transfer between parallel graphene sheets},\ }\href
  {https://doi.org/10.1103/PhysRevB.96.155437} {\bibfield  {journal} {\bibinfo
  {journal} {Phys. Rev. B}\ }\textbf {\bibinfo {volume} {96}},\ \bibinfo
  {pages} {155437} (\bibinfo {year} {2017})}\BibitemShut {NoStop}%
\bibitem [{\citenamefont {Tang}\ and\ \citenamefont {Wang}(2018)}]{GT18}%
  \BibitemOpen
  \bibfield  {author} {\bibinfo {author} {\bibfnamefont {G.}~\bibnamefont
  {Tang}}\ and\ \bibinfo {author} {\bibfnamefont {J.-S.}\ \bibnamefont
  {Wang}},\ }\bibfield  {title} {\bibinfo {title} {Heat transfer statistics in
  extreme-near-field radiation},\ }\href
  {https://doi.org/10.1103/PhysRevB.98.125401} {\bibfield  {journal} {\bibinfo
  {journal} {Phys. Rev. B}\ }\textbf {\bibinfo {volume} {98}},\ \bibinfo
  {pages} {125401} (\bibinfo {year} {2018})}\BibitemShut {NoStop}%
\bibitem [{\citenamefont {Tang}\ \emph {et~al.}(2019)\citenamefont {Tang},
  \citenamefont {Yap}, \citenamefont {Ren},\ and\ \citenamefont {Wang}}]{GT19}%
  \BibitemOpen
  \bibfield  {author} {\bibinfo {author} {\bibfnamefont {G.}~\bibnamefont
  {Tang}}, \bibinfo {author} {\bibfnamefont {H.~H.}\ \bibnamefont {Yap}},
  \bibinfo {author} {\bibfnamefont {J.}~\bibnamefont {Ren}},\ and\ \bibinfo
  {author} {\bibfnamefont {J.-S.}\ \bibnamefont {Wang}},\ }\bibfield  {title}
  {\bibinfo {title} {Anomalous near-field heat transfer in carbon-based
  nanostructures with edge states},\ }\href
  {https://doi.org/10.1103/PhysRevApplied.11.031004} {\bibfield  {journal}
  {\bibinfo  {journal} {Phys. Rev. Appl.}\ }\textbf {\bibinfo {volume} {11}},\
  \bibinfo {pages} {031004} (\bibinfo {year} {2019})}\BibitemShut {NoStop}%
\bibitem [{\citenamefont {Wise}\ \emph {et~al.}(2022)\citenamefont {Wise},
  \citenamefont {Roubinowitz}, \citenamefont {Belzig},\ and\ \citenamefont
  {Basko}}]{Wise22}%
  \BibitemOpen
  \bibfield  {author} {\bibinfo {author} {\bibfnamefont {J.~L.}\ \bibnamefont
  {Wise}}, \bibinfo {author} {\bibfnamefont {N.}~\bibnamefont {Roubinowitz}},
  \bibinfo {author} {\bibfnamefont {W.}~\bibnamefont {Belzig}},\ and\ \bibinfo
  {author} {\bibfnamefont {D.~M.}\ \bibnamefont {Basko}},\ }\bibfield  {title}
  {\bibinfo {title} {Signature of resonant modes in radiative heat current
  noise spectrum},\ }\href {https://doi.org/10.1103/PhysRevB.106.165407}
  {\bibfield  {journal} {\bibinfo  {journal} {Phys. Rev. B}\ }\textbf {\bibinfo
  {volume} {106}},\ \bibinfo {pages} {165407} (\bibinfo {year}
  {2022})}\BibitemShut {NoStop}%
\bibitem [{\citenamefont {Wang}\ \emph {et~al.}(2023)\citenamefont {Wang},
  \citenamefont {Peng}, \citenamefont {Zhang}, \citenamefont {Zhang},\ and\
  \citenamefont {Zhu}}]{JSW23}%
  \BibitemOpen
  \bibfield  {author} {\bibinfo {author} {\bibfnamefont {J.-S.}\ \bibnamefont
  {Wang}}, \bibinfo {author} {\bibfnamefont {J.}~\bibnamefont {Peng}}, \bibinfo
  {author} {\bibfnamefont {Z.-Q.}\ \bibnamefont {Zhang}}, \bibinfo {author}
  {\bibfnamefont {Y.-M.}\ \bibnamefont {Zhang}},\ and\ \bibinfo {author}
  {\bibfnamefont {T.}~\bibnamefont {Zhu}},\ }\bibfield  {title} {\bibinfo
  {title} {Transport in electron-photon systems},\ }\href@noop {} {\bibfield
  {journal} {\bibinfo  {journal} {Front. Phys.}\ }\textbf {\bibinfo {volume}
  {18}},\ \bibinfo {pages} {43602} (\bibinfo {year} {2023})}\BibitemShut
  {NoStop}%
\bibitem [{\citenamefont {Wang}\ and\ \citenamefont {Antezza}(2023)}]{JSW23_2}%
  \BibitemOpen
  \bibfield  {author} {\bibinfo {author} {\bibfnamefont {J.-S.}\ \bibnamefont
  {Wang}}\ and\ \bibinfo {author} {\bibfnamefont {M.}~\bibnamefont {Antezza}},\
  }\href@noop {} {\bibinfo {title} {Photon mediated energy, linear and angular
  momentum transport in fullerene and graphene systems beyond local
  equilibrium}} (\bibinfo {year} {2023}),\ \Eprint
  {https://arxiv.org/abs/arXiv:2307.11361} {arXiv:2307.11361} \BibitemShut
  {NoStop}%
\bibitem [{\citenamefont {Aoki}\ \emph {et~al.}(2014)\citenamefont {Aoki},
  \citenamefont {Tsuji}, \citenamefont {Eckstein}, \citenamefont {Kollar},
  \citenamefont {Oka},\ and\ \citenamefont {Werner}}]{DMFT-RMP14}%
  \BibitemOpen
  \bibfield  {author} {\bibinfo {author} {\bibfnamefont {H.}~\bibnamefont
  {Aoki}}, \bibinfo {author} {\bibfnamefont {N.}~\bibnamefont {Tsuji}},
  \bibinfo {author} {\bibfnamefont {M.}~\bibnamefont {Eckstein}}, \bibinfo
  {author} {\bibfnamefont {M.}~\bibnamefont {Kollar}}, \bibinfo {author}
  {\bibfnamefont {T.}~\bibnamefont {Oka}},\ and\ \bibinfo {author}
  {\bibfnamefont {P.}~\bibnamefont {Werner}},\ }\bibfield  {title} {\bibinfo
  {title} {Nonequilibrium dynamical mean-field theory and its applications},\
  }\href {https://doi.org/10.1103/RevModPhys.86.779} {\bibfield  {journal}
  {\bibinfo  {journal} {Rev. Mod. Phys.}\ }\textbf {\bibinfo {volume} {86}},\
  \bibinfo {pages} {779} (\bibinfo {year} {2014})}\BibitemShut {NoStop}%
\bibitem [{\citenamefont {Li}\ \emph {et~al.}(2014)\citenamefont {Li},
  \citenamefont {Chen}, \citenamefont {Meng}, \citenamefont {Fang},
  \citenamefont {Xiao}, \citenamefont {Li}, \citenamefont {Hu}, \citenamefont
  {Xu}, \citenamefont {Tong}, \citenamefont {Wang}, \citenamefont {Liu},
  \citenamefont {Bao},\ and\ \citenamefont {Shen}}]{graphene-modulate-14}%
  \BibitemOpen
  \bibfield  {author} {\bibinfo {author} {\bibfnamefont {W.}~\bibnamefont
  {Li}}, \bibinfo {author} {\bibfnamefont {B.}~\bibnamefont {Chen}}, \bibinfo
  {author} {\bibfnamefont {C.}~\bibnamefont {Meng}}, \bibinfo {author}
  {\bibfnamefont {W.}~\bibnamefont {Fang}}, \bibinfo {author} {\bibfnamefont
  {Y.}~\bibnamefont {Xiao}}, \bibinfo {author} {\bibfnamefont {X.}~\bibnamefont
  {Li}}, \bibinfo {author} {\bibfnamefont {Z.}~\bibnamefont {Hu}}, \bibinfo
  {author} {\bibfnamefont {Y.}~\bibnamefont {Xu}}, \bibinfo {author}
  {\bibfnamefont {L.}~\bibnamefont {Tong}}, \bibinfo {author} {\bibfnamefont
  {H.}~\bibnamefont {Wang}}, \bibinfo {author} {\bibfnamefont {W.}~\bibnamefont
  {Liu}}, \bibinfo {author} {\bibfnamefont {J.}~\bibnamefont {Bao}},\ and\
  \bibinfo {author} {\bibfnamefont {Y.~R.}\ \bibnamefont {Shen}},\ }\bibfield
  {title} {\bibinfo {title} {Ultrafast all-optical graphene modulator},\ }\href
  {https://doi.org/10.1021/nl404356t} {\bibfield  {journal} {\bibinfo
  {journal} {Nano Lett.}\ }\textbf {\bibinfo {volume} {14}},\ \bibinfo {pages}
  {955} (\bibinfo {year} {2014})}\BibitemShut {NoStop}%
\bibitem [{\citenamefont {Tasolamprou}\ \emph {et~al.}(2019)\citenamefont
  {Tasolamprou}, \citenamefont {Koulouklidis}, \citenamefont {Daskalaki},
  \citenamefont {Mavidis}, \citenamefont {Kenanakis}, \citenamefont
  {Deligeorgis}, \citenamefont {Viskadourakis}, \citenamefont {Kuzhir},
  \citenamefont {Tzortzakis}, \citenamefont {Kafesaki}, \citenamefont
  {Economou},\ and\ \citenamefont {Soukoulis}}]{graphene-modulate-19}%
  \BibitemOpen
  \bibfield  {author} {\bibinfo {author} {\bibfnamefont {A.~C.}\ \bibnamefont
  {Tasolamprou}}, \bibinfo {author} {\bibfnamefont {A.~D.}\ \bibnamefont
  {Koulouklidis}}, \bibinfo {author} {\bibfnamefont {C.}~\bibnamefont
  {Daskalaki}}, \bibinfo {author} {\bibfnamefont {C.~P.}\ \bibnamefont
  {Mavidis}}, \bibinfo {author} {\bibfnamefont {G.}~\bibnamefont {Kenanakis}},
  \bibinfo {author} {\bibfnamefont {G.}~\bibnamefont {Deligeorgis}}, \bibinfo
  {author} {\bibfnamefont {Z.}~\bibnamefont {Viskadourakis}}, \bibinfo {author}
  {\bibfnamefont {P.}~\bibnamefont {Kuzhir}}, \bibinfo {author} {\bibfnamefont
  {S.}~\bibnamefont {Tzortzakis}}, \bibinfo {author} {\bibfnamefont
  {M.}~\bibnamefont {Kafesaki}}, \bibinfo {author} {\bibfnamefont {E.~N.}\
  \bibnamefont {Economou}},\ and\ \bibinfo {author} {\bibfnamefont {C.~M.}\
  \bibnamefont {Soukoulis}},\ }\bibfield  {title} {\bibinfo {title}
  {Experimental demonstration of ultrafast {TH}z modulation in a graphene-based
  thin film absorber through negative photoinduced conductivity},\ }\href
  {https://doi.org/10.1021/acsphotonics.8b01595} {\bibfield  {journal}
  {\bibinfo  {journal} {ACS Photonics}\ }\textbf {\bibinfo {volume} {6}},\
  \bibinfo {pages} {720} (\bibinfo {year} {2019})}\BibitemShut {NoStop}%
\bibitem [{\citenamefont {Mahan}(2017)}]{Mahan}%
  \BibitemOpen
  \bibfield  {author} {\bibinfo {author} {\bibfnamefont {G.~D.}\ \bibnamefont
  {Mahan}},\ }\bibfield  {title} {\bibinfo {title} {Tunneling of heat between
  metals},\ }\href {https://doi.org/10.1103/PhysRevB.95.115427} {\bibfield
  {journal} {\bibinfo  {journal} {Phys. Rev. B}\ }\textbf {\bibinfo {volume}
  {95}},\ \bibinfo {pages} {115427} (\bibinfo {year} {2017})}\BibitemShut
  {NoStop}%
\bibitem [{\citenamefont {Yu}\ \emph {et~al.}(2017)\citenamefont {Yu},
  \citenamefont {Manjavacas},\ and\ \citenamefont {Garc{\'i}a~de
  Abajo}}]{SPP-graphene3}%
  \BibitemOpen
  \bibfield  {author} {\bibinfo {author} {\bibfnamefont {R.}~\bibnamefont
  {Yu}}, \bibinfo {author} {\bibfnamefont {A.}~\bibnamefont {Manjavacas}},\
  and\ \bibinfo {author} {\bibfnamefont {F.~J.}\ \bibnamefont {Garc{\'i}a~de
  Abajo}},\ }\bibfield  {title} {\bibinfo {title} {Ultrafast radiative heat
  transfer},\ }\href {https://doi.org/10.1038/s41467-016-0013-x} {\bibfield
  {journal} {\bibinfo  {journal} {Nat. Commun.}\ }\textbf {\bibinfo {volume}
  {8}},\ \bibinfo {pages} {2} (\bibinfo {year} {2017})}\BibitemShut {NoStop}%
\bibitem [{\citenamefont {Wise}\ \emph {et~al.}(2020)\citenamefont {Wise},
  \citenamefont {Basko},\ and\ \citenamefont {Hekking}}]{plasmon20_1}%
  \BibitemOpen
  \bibfield  {author} {\bibinfo {author} {\bibfnamefont {J.~L.}\ \bibnamefont
  {Wise}}, \bibinfo {author} {\bibfnamefont {D.~M.}\ \bibnamefont {Basko}},\
  and\ \bibinfo {author} {\bibfnamefont {F.~W.~J.}\ \bibnamefont {Hekking}},\
  }\bibfield  {title} {\bibinfo {title} {Role of disorder in plasmon-assisted
  near-field heat transfer between two-dimensional metals},\ }\href
  {https://doi.org/10.1103/PhysRevB.101.205411} {\bibfield  {journal} {\bibinfo
   {journal} {Phys. Rev. B}\ }\textbf {\bibinfo {volume} {101}},\ \bibinfo
  {pages} {205411} (\bibinfo {year} {2020})}\BibitemShut {NoStop}%
\bibitem [{\citenamefont {Ying}\ and\ \citenamefont
  {Kamenev}(2020)}]{plasmon20_2}%
  \BibitemOpen
  \bibfield  {author} {\bibinfo {author} {\bibfnamefont {X.}~\bibnamefont
  {Ying}}\ and\ \bibinfo {author} {\bibfnamefont {A.}~\bibnamefont {Kamenev}},\
  }\bibfield  {title} {\bibinfo {title} {Plasmonic tuning of near-field heat
  transfer between graphene monolayers},\ }\href
  {https://doi.org/10.1103/PhysRevB.102.195426} {\bibfield  {journal} {\bibinfo
   {journal} {Phys. Rev. B}\ }\textbf {\bibinfo {volume} {102}},\ \bibinfo
  {pages} {195426} (\bibinfo {year} {2020})}\BibitemShut {NoStop}%
\bibitem [{\citenamefont {Chudnovskiy}\ \emph {et~al.}(2023)\citenamefont
  {Chudnovskiy}, \citenamefont {Levchenko},\ and\ \citenamefont
  {Kamenev}}]{Kamenev23}%
  \BibitemOpen
  \bibfield  {author} {\bibinfo {author} {\bibfnamefont {A.~L.}\ \bibnamefont
  {Chudnovskiy}}, \bibinfo {author} {\bibfnamefont {A.}~\bibnamefont
  {Levchenko}},\ and\ \bibinfo {author} {\bibfnamefont {A.}~\bibnamefont
  {Kamenev}},\ }\bibfield  {title} {\bibinfo {title} {Coulomb drag and heat
  transfer in strange metals},\ }\href
  {https://doi.org/10.1103/PhysRevLett.131.096501} {\bibfield  {journal}
  {\bibinfo  {journal} {Phys. Rev. Lett.}\ }\textbf {\bibinfo {volume} {131}},\
  \bibinfo {pages} {096501} (\bibinfo {year} {2023})}\BibitemShut {NoStop}%
\bibitem [{\citenamefont {Cohen-Tannoudji}\ \emph {et~al.}(1989)\citenamefont
  {Cohen-Tannoudji}, \citenamefont {Dupont-Roc},\ and\ \citenamefont
  {Grynberg}}]{QED}%
  \BibitemOpen
  \bibfield  {author} {\bibinfo {author} {\bibfnamefont {C.}~\bibnamefont
  {Cohen-Tannoudji}}, \bibinfo {author} {\bibfnamefont {J.}~\bibnamefont
  {Dupont-Roc}},\ and\ \bibinfo {author} {\bibfnamefont {G.}~\bibnamefont
  {Grynberg}},\ }\href@noop {} {\emph {\bibinfo {title} {Photons and Atoms:
  Introduction to Quantum Electrodynamics}}}\ (\bibinfo  {publisher} {Wiley,
  New York},\ \bibinfo {year} {1989})\BibitemShut {NoStop}%
\bibitem [{\citenamefont {Haug}\ and\ \citenamefont
  {Jauho}(2008)}]{Haug_Jauho}%
  \BibitemOpen
  \bibfield  {author} {\bibinfo {author} {\bibfnamefont {H.}~\bibnamefont
  {Haug}}\ and\ \bibinfo {author} {\bibfnamefont {A.-P.}\ \bibnamefont
  {Jauho}},\ }\href@noop {} {\emph {\bibinfo {title} {Quantum kinetics in
  transport and optics of semiconductors}}},\ Vol.~\bibinfo {volume} {2}\
  (\bibinfo  {publisher} {Springer},\ \bibinfo {year} {2008})\BibitemShut
  {NoStop}%
\bibitem [{\citenamefont {Chen}\ \emph
  {et~al.}(2015{\natexlab{a}})\citenamefont {Chen}, \citenamefont {ShangGuan},\
  and\ \citenamefont {Wang}}]{Chen_2015}%
  \BibitemOpen
  \bibfield  {author} {\bibinfo {author} {\bibfnamefont {J.}~\bibnamefont
  {Chen}}, \bibinfo {author} {\bibfnamefont {M.}~\bibnamefont {ShangGuan}},\
  and\ \bibinfo {author} {\bibfnamefont {J.}~\bibnamefont {Wang}},\ }\bibfield
  {title} {\bibinfo {title} {A gauge invariant theory for time dependent heat
  current},\ }\href {https://doi.org/10.1088/1367-2630/17/5/053034} {\bibfield
  {journal} {\bibinfo  {journal} {New J. Phys}\ }\textbf {\bibinfo {volume}
  {17}},\ \bibinfo {pages} {053034} (\bibinfo {year}
  {2015}{\natexlab{a}})}\BibitemShut {NoStop}%
\bibitem [{\citenamefont {Langreth}\ and\ \citenamefont
  {Nordlander}(1991)}]{Langreth}%
  \BibitemOpen
  \bibfield  {author} {\bibinfo {author} {\bibfnamefont {D.~C.}\ \bibnamefont
  {Langreth}}\ and\ \bibinfo {author} {\bibfnamefont {P.}~\bibnamefont
  {Nordlander}},\ }\bibfield  {title} {\bibinfo {title} {Derivation of a master
  equation for charge-transfer processes in atom-surface collisions},\ }\href
  {https://doi.org/10.1103/PhysRevB.43.2541} {\bibfield  {journal} {\bibinfo
  {journal} {Phys. Rev. B}\ }\textbf {\bibinfo {volume} {43}},\ \bibinfo
  {pages} {2541} (\bibinfo {year} {1991})}\BibitemShut {NoStop}%
\bibitem [{\citenamefont {Meir}\ and\ \citenamefont
  {Wingreen}(1992)}]{Wingreen92}%
  \BibitemOpen
  \bibfield  {author} {\bibinfo {author} {\bibfnamefont {Y.}~\bibnamefont
  {Meir}}\ and\ \bibinfo {author} {\bibfnamefont {N.~S.}\ \bibnamefont
  {Wingreen}},\ }\bibfield  {title} {\bibinfo {title} {Landauer formula for the
  current through an interacting electron region},\ }\href
  {https://doi.org/10.1103/PhysRevLett.68.2512} {\bibfield  {journal} {\bibinfo
   {journal} {Phys. Rev. Lett.}\ }\textbf {\bibinfo {volume} {68}},\ \bibinfo
  {pages} {2512} (\bibinfo {year} {1992})}\BibitemShut {NoStop}%
\bibitem [{\citenamefont {B\"uttiker}\ \emph {et~al.}(1993)\citenamefont
  {B\"uttiker}, \citenamefont {Pr\^etre},\ and\ \citenamefont
  {Thomas}}]{Buttiker93}%
  \BibitemOpen
  \bibfield  {author} {\bibinfo {author} {\bibfnamefont {M.}~\bibnamefont
  {B\"uttiker}}, \bibinfo {author} {\bibfnamefont {A.}~\bibnamefont
  {Pr\^etre}},\ and\ \bibinfo {author} {\bibfnamefont {H.}~\bibnamefont
  {Thomas}},\ }\bibfield  {title} {\bibinfo {title} {Admittance of small
  conductors},\ }\href {https://doi.org/10.1103/PhysRevLett.71.465} {\bibfield
  {journal} {\bibinfo  {journal} {Phys. Rev. Lett.}\ }\textbf {\bibinfo
  {volume} {71}},\ \bibinfo {pages} {465} (\bibinfo {year} {1993})}\BibitemShut
  {NoStop}%
\bibitem [{\citenamefont {Wang}\ \emph {et~al.}(1999)\citenamefont {Wang},
  \citenamefont {Wang},\ and\ \citenamefont {Guo}}]{Wang99}%
  \BibitemOpen
  \bibfield  {author} {\bibinfo {author} {\bibfnamefont {B.}~\bibnamefont
  {Wang}}, \bibinfo {author} {\bibfnamefont {J.}~\bibnamefont {Wang}},\ and\
  \bibinfo {author} {\bibfnamefont {H.}~\bibnamefont {Guo}},\ }\bibfield
  {title} {\bibinfo {title} {Current partition: A nonequilibrium {G}reen's
  function approach},\ }\href {https://doi.org/10.1103/PhysRevLett.82.398}
  {\bibfield  {journal} {\bibinfo  {journal} {Phys. Rev. Lett.}\ }\textbf
  {\bibinfo {volume} {82}},\ \bibinfo {pages} {398} (\bibinfo {year}
  {1999})}\BibitemShut {NoStop}%
\bibitem [{\citenamefont {Stefanucci}\ and\ \citenamefont {van
  Leeuwen}(2013)}]{Stefanucci}%
  \BibitemOpen
  \bibfield  {author} {\bibinfo {author} {\bibfnamefont {G.}~\bibnamefont
  {Stefanucci}}\ and\ \bibinfo {author} {\bibfnamefont {R.}~\bibnamefont {van
  Leeuwen}},\ }\href@noop {} {\emph {\bibinfo {title} {Nonequilibrium Many-Body
  Theory of Quantum Systems: A Modern Introduction}}}\ (\bibinfo  {publisher}
  {Cambridge University Press},\ \bibinfo {year} {2013})\BibitemShut {NoStop}%
\bibitem [{\citenamefont {Chen}\ \emph
  {et~al.}(2015{\natexlab{b}})\citenamefont {Chen}, \citenamefont {Santhanam},
  \citenamefont {Sandhu}, \citenamefont {Zhu},\ and\ \citenamefont
  {Fan}}]{photon_potential_15}%
  \BibitemOpen
  \bibfield  {author} {\bibinfo {author} {\bibfnamefont {K.}~\bibnamefont
  {Chen}}, \bibinfo {author} {\bibfnamefont {P.}~\bibnamefont {Santhanam}},
  \bibinfo {author} {\bibfnamefont {S.}~\bibnamefont {Sandhu}}, \bibinfo
  {author} {\bibfnamefont {L.}~\bibnamefont {Zhu}},\ and\ \bibinfo {author}
  {\bibfnamefont {S.}~\bibnamefont {Fan}},\ }\bibfield  {title} {\bibinfo
  {title} {Heat-flux control and solid-state cooling by regulating chemical
  potential of photons in near-field electromagnetic heat transfer},\ }\href
  {https://doi.org/10.1103/PhysRevB.91.134301} {\bibfield  {journal} {\bibinfo
  {journal} {Phys. Rev. B}\ }\textbf {\bibinfo {volume} {91}},\ \bibinfo
  {pages} {134301} (\bibinfo {year} {2015}{\natexlab{b}})}\BibitemShut
  {NoStop}%
\bibitem [{\citenamefont {Chen}\ \emph {et~al.}(2016)\citenamefont {Chen},
  \citenamefont {Santhanam},\ and\ \citenamefont {Fan}}]{photon_potential_16}%
  \BibitemOpen
  \bibfield  {author} {\bibinfo {author} {\bibfnamefont {K.}~\bibnamefont
  {Chen}}, \bibinfo {author} {\bibfnamefont {P.}~\bibnamefont {Santhanam}},\
  and\ \bibinfo {author} {\bibfnamefont {S.}~\bibnamefont {Fan}},\ }\bibfield
  {title} {\bibinfo {title} {Near-field enhanced negative luminescent
  refrigeration},\ }\href {https://doi.org/10.1103/PhysRevApplied.6.024014}
  {\bibfield  {journal} {\bibinfo  {journal} {Phys. Rev. Appl.}\ }\textbf
  {\bibinfo {volume} {6}},\ \bibinfo {pages} {024014} (\bibinfo {year}
  {2016})}\BibitemShut {NoStop}%
\bibitem [{\citenamefont {Zhu}\ \emph {et~al.}(2019)\citenamefont {Zhu},
  \citenamefont {Fiorino}, \citenamefont {Thompson}, \citenamefont
  {Mittapally}, \citenamefont {Meyhofer},\ and\ \citenamefont
  {Reddy}}]{photon_potential_19}%
  \BibitemOpen
  \bibfield  {author} {\bibinfo {author} {\bibfnamefont {L.}~\bibnamefont
  {Zhu}}, \bibinfo {author} {\bibfnamefont {A.}~\bibnamefont {Fiorino}},
  \bibinfo {author} {\bibfnamefont {D.}~\bibnamefont {Thompson}}, \bibinfo
  {author} {\bibfnamefont {R.}~\bibnamefont {Mittapally}}, \bibinfo {author}
  {\bibfnamefont {E.}~\bibnamefont {Meyhofer}},\ and\ \bibinfo {author}
  {\bibfnamefont {P.}~\bibnamefont {Reddy}},\ }\bibfield  {title} {\bibinfo
  {title} {Near-field photonic cooling through control of the chemical
  potential of photons},\ }\href {https://doi.org/10.1038/s41586-019-0918-8}
  {\bibfield  {journal} {\bibinfo  {journal} {Nature}\ }\textbf {\bibinfo
  {volume} {566}},\ \bibinfo {pages} {239} (\bibinfo {year}
  {2019})}\BibitemShut {NoStop}%
\bibitem [{\citenamefont {Tsuji}\ \emph {et~al.}(2008)\citenamefont {Tsuji},
  \citenamefont {Oka},\ and\ \citenamefont {Aoki}}]{Tsuji08}%
  \BibitemOpen
  \bibfield  {author} {\bibinfo {author} {\bibfnamefont {N.}~\bibnamefont
  {Tsuji}}, \bibinfo {author} {\bibfnamefont {T.}~\bibnamefont {Oka}},\ and\
  \bibinfo {author} {\bibfnamefont {H.}~\bibnamefont {Aoki}},\ }\bibfield
  {title} {\bibinfo {title} {Correlated electron systems periodically driven
  out of equilibrium: $\text{Floquet}+\text{DMFT}$ formalism},\ }\href
  {https://doi.org/10.1103/PhysRevB.78.235124} {\bibfield  {journal} {\bibinfo
  {journal} {Phys. Rev. B}\ }\textbf {\bibinfo {volume} {78}},\ \bibinfo
  {pages} {235124} (\bibinfo {year} {2008})}\BibitemShut {NoStop}%
\bibitem [{\citenamefont {Honeychurch}\ and\ \citenamefont
  {Kosov}(2023{\natexlab{a}})}]{Kosov23_1}%
  \BibitemOpen
  \bibfield  {author} {\bibinfo {author} {\bibfnamefont {T.~D.}\ \bibnamefont
  {Honeychurch}}\ and\ \bibinfo {author} {\bibfnamefont {D.~S.}\ \bibnamefont
  {Kosov}},\ }\bibfield  {title} {\bibinfo {title} {Quantum transport in driven
  systems with vibrations: Floquet nonequilibrium {G}reen's functions and the
  self-consistent {B}orn approximation},\ }\href
  {https://doi.org/10.1103/PhysRevB.107.035410} {\bibfield  {journal} {\bibinfo
   {journal} {Phys. Rev. B}\ }\textbf {\bibinfo {volume} {107}},\ \bibinfo
  {pages} {035410} (\bibinfo {year} {2023}{\natexlab{a}})}\BibitemShut
  {NoStop}%
\bibitem [{\citenamefont {Honeychurch}\ and\ \citenamefont
  {Kosov}(2023{\natexlab{b}})}]{Kosov23_2}%
  \BibitemOpen
  \bibfield  {author} {\bibinfo {author} {\bibfnamefont {T.~D.}\ \bibnamefont
  {Honeychurch}}\ and\ \bibinfo {author} {\bibfnamefont {D.~S.}\ \bibnamefont
  {Kosov}},\ }\href@noop {} {\bibinfo {title} {Floquet nonequilibrium {G}reen's
  functions with fluctuation-exchange approximation: Application to
  periodically driven capacitively coupled quantum dots}} (\bibinfo {year}
  {2023}{\natexlab{b}}),\ \Eprint {https://arxiv.org/abs/arXiv:2307.09774}
  {arXiv:2307.09774} \BibitemShut {NoStop}%
\bibitem [{\citenamefont {Ludovico}\ \emph {et~al.}(2014)\citenamefont
  {Ludovico}, \citenamefont {Lim}, \citenamefont {Moskalets}, \citenamefont
  {Arrachea},\ and\ \citenamefont {S\'anchez}}]{Sanchez14}%
  \BibitemOpen
  \bibfield  {author} {\bibinfo {author} {\bibfnamefont {M.~F.}\ \bibnamefont
  {Ludovico}}, \bibinfo {author} {\bibfnamefont {J.~S.}\ \bibnamefont {Lim}},
  \bibinfo {author} {\bibfnamefont {M.}~\bibnamefont {Moskalets}}, \bibinfo
  {author} {\bibfnamefont {L.}~\bibnamefont {Arrachea}},\ and\ \bibinfo
  {author} {\bibfnamefont {D.}~\bibnamefont {S\'anchez}},\ }\bibfield  {title}
  {\bibinfo {title} {Dynamical energy transfer in ac-driven quantum systems},\
  }\href {https://doi.org/10.1103/PhysRevB.89.161306} {\bibfield  {journal}
  {\bibinfo  {journal} {Phys. Rev. B}\ }\textbf {\bibinfo {volume} {89}},\
  \bibinfo {pages} {161306} (\bibinfo {year} {2014})}\BibitemShut {NoStop}%
\end{thebibliography}%

\end{document}